\documentclass[twocolumn, showpacs, english, preprintnumbers, amsmath, amssymb, superscriptaddress, aps, prl,longbibliography]{revtex4-2}
\usepackage{silence}
\usepackage{xspace}
\newcommand{\MATLAB}{\textsc{Matlab}\xspace}
\usepackage[subpreambles=true]{standalone}
\usepackage[utf8]{inputenc}
\usepackage{graphicx}
\usepackage{grffile}
\usepackage[usenames,dvipsnames]{xcolor}
\usepackage{amsmath,bm}
\usepackage[normalem]{ulem}
\usepackage{appendix} 
\definecolor{orange}{rgb}{1,0.5,0}
\definecolor{goodgreen}{rgb}{0.1,0.5,0}
\definecolor{goodred}{rgb}{0.7,0,0}

\renewcommand\vec{\boldsymbol}

\usepackage{tikz}
\makeatletter

\usepackage[colorlinks,urlcolor=goodgreen,citecolor=blue,linkcolor=goodred]{hyperref}

\newcommand{\orcid}[1]{\href{https://orcid.org/#1}{\includegraphics[width=8pt]{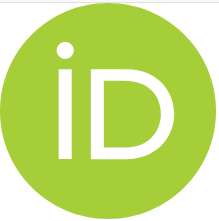}}}

\usepackage{color}

\usepackage{float}

\let\oldepsilon\epsilon \let\epsilon\varepsilon \let\varepsilon\oldepsilon

\makeatother

\usepackage{babel}

\makeatother

\begin{document}
\title{ 
A universal phenomenology of charge-spin interconversion and dynamics in diffusive systems with spin-orbit coupling.}

\author{Tim Kokkeler\orcid{0000-0001-8681-3376}}
\email{tim.kokkeler@dipc.org}

\affiliation{Donostia International Physics Center (DIPC), 20018 Donostia--San
Sebastián, Spain}
\affiliation{University of Twente, 7522 NB Enschede, The Netherlands}

\author{F. Sebastian Bergeret\orcid{0000-0001-6007-4878}}
\email{fs.bergeret@csic.es}

\affiliation{Centro de Física de Materiales (CFM-MPC) Centro Mixto CSIC-UPV/EHU,
E-20018 Donostia-San Sebastián, Spain}
\affiliation{Donostia International Physics Center (DIPC), 20018 Donostia--San
Sebastián, Spain}

\author{I. V. Tokatly\orcid{0000-0001-6288-0689}}
\email{ilya.tokatly@ehu.es}

\affiliation{Donostia International Physics Center (DIPC), 20018 Donostia--San
Sebastián, Spain}
\affiliation{Departamento de Polimeros y Materiales
Avanzados, Universidad del Pais Vasco UPV/EHU, 20018 Donostia-San Sebastian,
Spain}
\affiliation{IKERBASQUE, Basque Foundation for Science, 48009 Bilbao, Spain}

\begin{abstract}
We present a unified description of transport  %present an effective field theory  for a unified description of transport 
in   normal and superconducting metals in the presence of generic spin-orbit coupling (SOC). %We present an effective field theory  for a unified description of transport in   normal and superconducting metals in the presence of generic spin-orbit coupling (SOC).  
The structure of the quantum kinetic theory in the diffusive regime is determined by a set of fundamental constraints -- charge conjugation symmetry, the causality principle, and the crystal symmetry of a material. These symmetries uniquely fix the action of the Keldysh nonlinear $\sigma$ model (NLSM), which at the saddle point yields the quantum kinetic Usadel-type equation, the equation that describes the main transport features of a system. Our phenomenological approach is reminiscent of the Ginzburg-Landau theory, but is valid for superconductors in the whole temperature range,  describes the diffusive transport in the normal state, and naturally captures the effects of superconducting fluctuations. As an application, we derive the NLSM and the corresponding quantum transport equations which include all effects of spin-orbit coupling, allowed by the crystal symmetry, for example, the spin Hall, spin current swapping or spin-galvanic effects. Our approach can be extended to derive transport equations in systems with broken time reversal symmetry, as well as to the description of hybrid interfaces, where the spin-charge interconversion can be enhanced due to strong interfacial SOC.  
\end{abstract}
\maketitle

Electronic transport in materials with spin-orbit coupling (SOC) has attracted a lot of interest in recent decades, particularly due to the possibility of spin-charge interconversion  \cite{sinova2015spin,ganichev2019spin}. This interest has extended to systems where superconductivity coexists with spin-orbit coupling, motivated by intensive research in  the fields of topological superconductivity \cite{sato2017topological}, two-dimensional superconductors and van der Waals heterostructures \cite{novoselov20162d}, and noncentrosymmetric superconductors  \cite{smidman2017superconductivity}. The interplay between superconductivity and spin-orbit coupling underlies the physics of superconducting magnetoelectric and spin diode effects 
\cite{nadeem2023superconducting,bobkova2022magnetoelectric}.

Real devices used for experimental exploration of transport properties typically feature mesoscopic sizes, where disorder is usually unavoidable.  This underscores the importance of deriving effective low-energy quantum kinetic theories in the diffusive regime. These theories include the powerful quasiclassical approach,  which has been used to describe the interplay between superconductivity and spin-orbit coupling (SOC) in different systems \cite{ilic2020unified,houzet2015quasiclassical,bobkova2017quasiclassical,bergeret2014spin,bergeret2016manifestation,tokatly2017usadel,bergeret2015theory,virtanen2021magnetoelectric,virtanen2022nonlinear,bobkov2023magnetic,bobkova2016magnetoelectrics,bobkova2022magnetoelectric,bobkova2017gauge,amundsen2019quasiclassical,linder2022quasiclassical,linder2011spin}.

The central kinetic equation in diffusive superconductors is the Usadel equation \cite{usadel1970generalized}. In systems without spin-orbit coupling, it can be derived directly from the Gorkov equations following a standard procedure \cite{larkin1986,belzig1999quasiclassical}. However, mixing of charge and spin degrees of freedom in the presence of small, but finite spin-orbit coupling (SOC) prevents a straightforward application of such a standard scheme. Recently, it has been realized \cite{virtanen2021magnetoelectric,virtanen2022nonlinear} that in this case, the Usadel equation can  be conveniently  derived from the nonlinear sigma model (NLSM)  \cite{efetov1980interaction,wegner1979mobility,finkelshtein1983effect}, as a saddle point of effective action for disordered conductors. In Refs.~\cite{virtanen2021magnetoelectric,virtanen2022nonlinear} the NLSM was derived for the special cases of extrinsic and Rashba-like intrinsic SOC, following the standard microscopic procedure of averaging over Gaussian disorder (see, for example, Refs.~\cite{EfetovBook,lerner2003nonlinear,altland2010condensed,kamenev2023field} for different versions of the NLSM).  Apparently, such an approach can only be effective for simplified models, while for realistic materials it becomes prohibitively tedious because of their complex spin-dependent electronic structure and the necessity to integrate various mechanisms of SOC.

Here we propose another approach to derive the corresponding quantum kinetic equations via the NLSM, that, similarly to the Ginzburg-Landau (GL) theory, relies solely on symmetry arguments. This allows us to incorporate effects of spin-orbit coupling into the NLSM and formulate a phenomenological quantum kinetic theory. Our theory captures all symmetry-allowed transport effects associated with spin, charge, and their coupling, such as spin-Hall and spin-galvanic effects. In contrast to GL theory,  this phenomenology is valid for arbitrary temperatures,  away from equilibrium, and covers the normal state. In the following, we identify the  basic symmetries and explain how to  construct the generalized NLSM. Then, the saddle point condition leads us to our main objective: obtaining the  general quantum kinetic equation for disordered conductors with spin-orbit coupling, which reveals all physical effects associated with the different terms of the action. 
%\textcolor{red}{The obtained quantum kinetic equation can be applied to a broad range of materials with a generic spin-dependent electronic structure, including various types of spin-orbit coupling, and allowing for the simultaneous presence of spin-Hall and magneto-toroidic tensors. Our model recovers the equations found in the two specific cases most studied in the literature: SU(2) covariant linear-in-momentum spin-orbit coupling \cite{virtanen2022nonlinear} and isotropic extrinsic SOC \cite{virtanen2021magnetoelectric}. }

The obtained quantum kinetic equation can be used for a broad scope of materials. In the specific cases of an SU(2) covariant Rashba type spin-orbit coupling or exclusively isotropic extrinsic SOC our model recovers the equations found in \cite{virtanen2022nonlinear} and \cite{virtanen2021magnetoelectric} respectively. However, the presented equations have a far more general scope and can be used as well for materials with different types of spin-orbit coupling and the combination of several sources of SOC. %{\color{magenta} [I THINK THE TEXT IN BLUE SUBSTITUTE THE TEXT IN RED RIGHT?]} \textcolor{blue}{\bf[IT: Red are the revisions made on the 1st round and blue are the present modifications made in the response to the 2nd referee.]}

Like for the G-L theory, we develop our theory not from a microscopic Hamiltonian, but by directly constructing an effective action based on the relevant symmetries.
Specifically, we derive the Keldysh contour version of NLSM \cite{kamenev2009keldysh,kamenev2023field,levchenko2009transport,feigelman2000keldysh} because of its flexibility in both the equilibrium and nonequilibrium setting \cite{van2006introduction}. In thermal equilibrium, the saddle point value of NLSM action can be connected to the quasiclassical Luttinger-Ward free energy functional \cite{virtanen2020quasiclassical,virtanen2024numerical}.
Formally, the NLSM is defined by the action functional $S[Q,\{A\}]$ which depends on a matrix field $Q$, subject to the constraint $Q^2=1$, and a set $\{A\}$ of external fields such as the exchange field and vector or scalar potentials.  In the Keldysh contour representation, the basic field $Q(\mathbf{r},t,t')$ is a 4$\times$4 matrix in the spin$\otimes$Nambu space \footnote{By the Nambu space we mean the space composed of the physical electronic states and their counterparts obtained by the time-reversal operation.} \cite{Skvotsov2000,kamenev2023field}, with time arguments $t$ and $t'$ placed on the two-branch (forward and backward) Keldysh contour. In practice, the Keldysh matrix representation is typically adopted, in which $t$ and $t'$ belong to the physical time axis, while the contour branches appear as an additional matrix index so that $Q$ becomes an 8$\times$8 matrix. Physical observables conjugated to external fields, and correlation functions are obtained by differentiating the generating functional $Z = \int [dQ] e^{iS[Q,\{A\}]}$. The expectation value of the $Q$-field, $g(t,t') = \int [dQ] Q(t,t') e^{iS[Q]}=\langle Q(t,t')\rangle$, gives the contour time-ordered quasiclassical Green function, providing direct access to one-particle spectral properties. Moreover, this implies that at the saddle point level, the NLSM generates transport equations for the quasiclassical Green's function $g$ \cite{kamenev2023field}.   Obtaining  these kinetic equations for $g$ in superconductors with generic types of spin-charge coupling is, in fact, the main objective of the present work.

It is instructive to begin by analyzing the microscopically derived NLSM for an isotropic superconductor in the presence of extrinsic SOC (spin-orbit scattering at a disorder potential) \cite{virtanen2021magnetoelectric}. The corresponding action reads as follows,
\begin{align}
&iS[Q] =\frac{\pi\nu}{2}\text{Tr}\Big(\hat{\omega}\tau_{3}Q + \hat{\Delta}Q + i\mathbf{h}{\bm\sigma}\tau_{3}Q -\frac{D}{4}(\nabla Q)^{2}\nonumber \\
- & \frac{D}{4}\epsilon_{ijk}\big[\theta\sigma_{i}Q\partial_{j}Q\partial_{k}Q-i\varkappa\sigma_i\partial_{j}Q\partial_{k}Q\big]+\frac{\Gamma}{8}\sigma_{k}Q\sigma_{k}Q\Big),
\label{eq:action0}
\end{align}
where the Pauli matrices $\sigma_i$ and $\tau_i$ span the spin and the Nambu spaces, respectively, $\hat{\omega}_{tt'}=\partial_t\delta(t-t')$ is the time derivative operator, $D$ is the diffusion coefficient, and $\hat{\Delta}=\tau_2|\Delta|e^{i\tau_3\varphi}$ is the superconducting order parameter matrix which pairs time-reversed partners \cite{anderson1959theory}. The second line in Eq.~\eqref{eq:action0} describes the effects of extrinsic SOC: the spin Hall effect, which connects mutually orthogonal charge and spin currents, the spin current swapping  \cite{lifshits2009swapping}, which mixes spin currents with "swapped" directions of flow and spin, and the spin relaxation. These are parameterized by the spin Hall angle $\theta$,  the swapping coefficient $\varkappa$, and the spin relaxation rate $\Gamma$,  respectively. The external electromagnetic potential is introduced through the replacement $\partial_{\mu}\mapsto\partial_{\mu}-i[A_{\mu}\tau_{3},\cdot]$, and the Zeeman or exchange field is denoted by $\mathbf{h}$. 

In equilibrium, this action can be related to the free energy $F[g]$ considered as a functional of the quasiclassical Green's function \cite{virtanen2020quasiclassical}. At a given temperature $T$, the mapping $TS[Q]\mapsto iF[g]$ is achieved by replacing $\hat{\omega}_{tt'}\rightarrow\omega_{n}$, the Matsubara frequency, and $Q\rightarrow g$. This connection becomes especially useful for superconductors as the minimization of $F[g]$ gives the equilibrium Usadel equation and determines  the thermodynamics of the system. 

%As indicated before, this action is valid both in and out of equilibrium. In equilibrium one can substitute $\hat{\omega}_{tt'}\xrightarrow{}\omega_{n}$ following SM Sec. III, where $\omega_{n} = (2n+1)\pi T$ are the Matsubara frequencies, $n$ is an integer and $T$ is temperature. Accordingly, we need to replace time integration by summation over these frequencies. In this limit the resulting action is related by a constant to the free energy of the system \cite{virtanen2020quasiclassical,virtanen2024numerical}. Out of equilibrium expression is not possible. However, if there is no explicit time dependence, we may impose time translation invariance, and therefore, using a Fourier transform on $t-t'$, substitute $\hat{\omega}_{tt'}\xrightarrow{}-iE$ and replace time integration with energy integration. With this the energy spectrum resolved properties of the system can be determined. For out of equilibrium systems with explicit time dependence, the full structure with two time arguments in Eq. (\ref{eq:action0}) is required.

By inspecting the structure of $S[Q]$ in Eq.~\eqref{eq:action0}, we can observe the following. In general, the action represents a low-energy gradient expansion combined with an expansion in powers of spin Pauli matrices $\sigma_i$, and involves $\tau_3$ in the terms/fields breaking the time-reversal symmetry. Each term in $S$ is a scalar composed of $Q$, $\partial_k Q$, and $\sigma_i$, and respecting the spatial symmetry of the system. In fact, the terms present in Eq.~\eqref{eq:action0} exhaust all scalar rotation invariant forms up to the second order in $\partial_k Q$ and $\sigma$-matrices, under the constraint $Q^2=1$. Therefore, the structure of all terms in the action, including those attributed to SOC, can be reconstructed from the space-time symmetry arguments. However, such arguments still allow for arbitrary complex coefficients in front of the invariants,  whereas the microscopically derived coefficients in Eq.~\eqref{eq:action0} are quite specific -- some of them are real, and some are purely imaginary. As we will see, this is controlled by another set of fundamental constraints unrelated to the space symmetry group of the system.

The first general symmetry of the NLSM is related to the redundancy of the Hilbert space in the Nambu description of superconductors, which involves both the full set of electronic states and their time-reversal conjugated counterparts. In nonsuperconducting systems, this technical doubling of the number of degrees of freedom is a convenient way to include fluctuations in the Cooperon channel. Because of the above redundancy, not all components of the matrix $Q$-field in NLSM are independent, but are constrained to satisfy the charge conjugation symmetry  \cite{efetov1980interaction,altland2010condensed,kamenev2023field}. In the Keldysh contour representation, this fundamental symmetry condition reads $Q=C{Q}^{T}C^{-1}$, where $C=-{i}\sigma_{2}\tau_{1}$ is the charge conjugation operator, and the transposition operation involves both the usual matrix transposition and the interchanging of the contour time arguments \footnote{It is worth noting that the charge conjugation is an exact and universal symmetry of the Bogolyubov-de Gennes equations, which is the consequence of the redundancy of the Nambu formalism  \cite{gurarie2011single,ryu2010topological,schnyder2008classification}.}.
In the Keldysh matrix representation, after the standard Keldysh rotation \cite{kamenev2023field} the {\it  charge conjugation symmetry} of the 8$\times$8 matrix $Q$-field takes the form (see SM Sec. \ref{sec:ChargeConjugation} for details),
\begin{align}\label{eq:ArtificialDoubling}
Q=\rho_{1}\tau_{1}\sigma_{2}Q^{T}\sigma_{2}\tau_{1}\rho_{1} & ,
\end{align}
where $\rho_i$ are Pauli matrices in the Keldysh space \footnote{Physically, the charge conjugation symmetry can be understood as equivalence of describing dynamics of the system using electron or hole representations.}.

Another condition restricting the form of the action follows from the fact that the expectation value of $Q$ equals to the Keldysh contour quasiclassical Green's function $g(t,t')=\langle Q(t,t')\rangle$. On the one hand, Hermitian conjugation of a chronologically ordered propagator reverses the time ordering to anti-chronological. On the other hand, in the Keldysh matrix representation, the transformation between chronological and anti-chronological contour Green's functions can be realized as a matrix operation. This leads to the following fundamental symmetry condition of the action of NLSM (see SM Sec.\ref{sec:chronological}),
\begin{align}
iS[Q]=\big(iS[-\rho_{2}\tau_{3}Q^{\dagger}\tau_{3}\rho_{2}]\big)^{*}.
\label{eq:chronology}
\end{align}
In the following, we refer to this condition as the {\it  chronological or time-ordering symmetry}. It is worth noting that Eq.~\eqref{eq:chronology} ensures that all observables are real.

One can easily check that the action Eq.~\eqref{eq:action0} respects both symmetries  Eqs.~\eqref{eq:ArtificialDoubling}-\eqref{eq:chronology}, and that the condition Eq.~\eqref{eq:chronology} uniquely determines whether the coefficients are real or purely imaginary.  

In a more general case, the contribution to $iS[Q]$ of a given order in $\partial_kQ$ and $\sigma_k$ can be constructed as follows: 
(i) Construct all distinct scalar primitive forms which contain under the common trace the chosen number of $\partial_kQ$ and $\sigma_k$, along with all possible number of $Q$'s, consistent with the constraint $Q^2=1$ and the charge conjugation, Eq. \eqref{eq:ArtificialDoubling}; (ii) Determine the  phase of tensor coefficients of the primitives by imposing the chronological symmetry Eq.~\eqref{eq:chronology}; (iii)  The allowed tensor coefficients are determined by the crystal symmetry group.

To illustrate this procedure, we derive contributions to the NLSM which may appear in the presence of  SOC. More details are presented in SM Sec. \ref{sec:Mainterms}. Physically, SOC mixes the spin and orbital/translation degrees of freedom, resulting in the simultaneous presence of $\sigma$-matrices and spatial gradients in the NLSM.  We construct the nontrivial contributions with the lowest orders on both Pauli matrices and gradients and discuss their main physical effects.  Since SOC preserves the time-reversal, we only consider terms invariant under the operation  $Q\mapsto \tau_3\sigma_yQ^T\sigma_y\tau_3$. Moreover, we consider only the dominant local in time contribution. Incorporation of small time-nonlocal terms breaking the particle-hole symmetry in NLSM is discussed in \cite{schwietenonlinearsigma,Jitu2024}.

{ \it Second order -- spin precession and relaxation.}  In the second order, there are only three primitives consistent with the condition $Q^2=1$ and Eq.~\eqref{eq:ArtificialDoubling}, namely, ${\rm Tr}(D_{kj}\partial_kQ\partial_jQ)$,  ${\rm Tr}(\Gamma_{kj}\sigma_kQ\sigma_jQ)$, and ${\rm Tr}(\tilde{\alpha}_{kj}\sigma_kQ\partial_jQ)$. The chronological symmetry of Eq.~\eqref{eq:chronology} requires that $D_{kj}$ and $\Gamma_{kj}$ are real symmetric tensors, while $\Tilde{\alpha}_{kj}=i\alpha_{kj}$ is an imaginary pseudotensor. $D_{kj}$ and $\Gamma_{kj}$ correspond, respectively, to the diffusion coefficient and the spin relaxation rate, which can be anisotropic, depending on the crystal symmetry. A second rank pseudotensor $\alpha_{kj}$ is allowed only in gyrotropic materials \cite{AgranovichBook} and the corresponding term is responsible for the spin precession induced by SOC.  By parameterizing $\alpha_{jk}$ as follows, $\alpha_{jk} = \mathcal{A}_{jl}D_{lk}$ the second order part of the NLSM can be cast in a compact form of SU(2) covariant diffusion with an additional spin relaxation,
\begin{align}\label{eq:S2}
    iS_2 = \frac{\pi\nu}{8}{\rm Tr}\Big(-D_{kj}\tilde{\nabla}_kQ\tilde{\nabla}_jQ + 
    \frac{1}{4}\Tilde{\Gamma}_{kj}\sigma_kQ\sigma_jQ\Big)\; ,
\end{align}
where $\tilde{\nabla}_k = \partial_k -i[\hat{\mathcal{A}}_{k},\cdot]$ is the SU(2) covariant derivative with an effective gauge field $\hat{\mathcal{A}}_k=\mathcal{A}_{jk}\sigma_j$, and $\Tilde{\Gamma}_{kj}=\Gamma_{kj}-4D_{ab}\mathcal{A}_{ka}\mathcal{A}_{jb}$ \footnote{We note that stability of the system require $D_{kj}$ and $\Gamma_{kj}$ to be positive definite.}. The SU(2) covariant diffusion has been introduced microscopically \cite{bergeret2014spin,BerTok2013PRL} for a special class of $\mathbf{k}\cdot\mathbf{p}$ models with a Rashba-like linear SOC. Now we see that it is generic for gyrotropic materials. In realistic systems, the SU(2) symmetry of simple models is of course broken. However, at the leading order this leads only to extra spin relaxation, which can originate from the Dyakonov-Perel \cite{Dyakonov1971}, or Elliott-Yafet \cite{Yafet1952,elliot1954theory} mechanisms, or from the interplay of SOC with electron-electron interactions \cite{GlaIvc2004,MowVigTok2011PRB}.
%\footnote{\textcolor{blue}{In our approach we naturally recover the leading effect of interference of SOC with electron-electron interaction, which appears in form of spin relaxation. Possible modification of the standard time-nonlocal interaction terms \cite{finkelstein1983influence,chamon1999schwinger} is an interesting question for the future.}}.} 

{ \it Third order -- spin-charge coupling related terms}.  There are two types of third order terms:

{ \it (1) Terms linear in $\sigma$ and quadratic in gradients.} Two forms of this type are consistent with the charge conjugation constraint, ${\rm Tr}\big(\tilde{\theta}_{ijk}\sigma_iQ\partial_{j}Q\partial_{k}Q\big)$ and  ${\rm Tr}\big(\tilde{\varkappa}_{ijk}\sigma_i\partial_{j}Q\partial_{k}Q\big)$, where the coefficients must be pseudotensors (to ensure that the forms are scalars) which are antisymmetric in the last couple of indices, $\Tilde{\theta}_{ijk}=-\Tilde{\theta}_{ikj}$, and $\Tilde{\varkappa}_{ijk}=-\Tilde{\varkappa}_{ikj}$. Finally, the symmetry Eq.~\eqref{eq:chronology} yields, $\Tilde{\theta}_{ijk}=\Tilde{\theta}_{ijk}^*$, and $\Tilde{\varkappa}_{ijk}=-\Tilde{\varkappa}_{ijk}^*$. By parameterizing the above third rank tensors in terms of dual second rank tensors, $\theta_{jk}$ and $\varkappa_{jk}$, we represent the type (1) contribution in the form
\begin{align} \label{S-SHE}
    iS_{3}^{(1)} = \frac{\pi\nu}{8}{\rm Tr}\Big(D\theta_{il}\epsilon_{ljk} \sigma_iQ\partial_{j}Q\partial_{k}Q - i D\varkappa_{il}\epsilon_{ljk} \sigma_i\partial_{j}Q\partial_{k}Q\Big),
\end{align}
which is a generalization of the  SHE and the spin swapping  terms for  anisotropic systems \cite{virtanen2021magnetoelectric}, see Eq.~\eqref{eq:action0}. 

{ \it (2) Terms quadratic in $\sigma$ and linear in gradients.} Along the same line of arguments, we identify two allowed scalar forms ${\rm Tr}\big(\tilde{\gamma}_{ijk} \sigma_i 
Q\sigma_{j}Q\partial_{k}Q\big)$ and ${\rm Tr}\big(\beta_{ijk} \sigma_i Q
\sigma_{j}\partial_{k}Q\big)$, where $\tilde{\gamma}_{ijk}=-\tilde{\gamma}_{jik}$ is purely imaginary, and $\beta_{ijk}=\beta_{jik}$ is real. The corresponding contribution to the action can thus be represented as, 
\begin{align}
    \label{S-SGE}
    iS_{3}^{(2)} = \frac{\pi\nu}{16}{\rm Tr}\Big(-i\epsilon_{ijl}\gamma_{lk} \sigma_iQ\sigma_{j}Q\partial_{k}Q + \beta_{ijk} \sigma_i Q 
\sigma_{j}\partial_{k}Q\Big)\; ,
\end{align}
where $i\gamma_{lj}$ is a pseudotensor dual to the tensor $\tilde{\gamma}_{ijk}$. Either term in Eq.~\eqref{S-SGE} requires breaking of inversion. The first one is allowed in the 18 gyrotropic classes, whereas the second term, containing a third rank tensor symmetric in the
first pair of indexes, may exist in the 20 piezoelectric crystal classes. We notice that the second term in Eq.~\eqref{S-SGE} can be written as a total derivative
${\rm Tr}\big(\frac{1}{2}\beta_{ijk}\partial_{k}[\sigma_{i}Q\sigma_{j}Q]\big)$, and therefore it does not contribute to the bulk action. For inhomogeneous $\beta_{ijk}$, it locally corrects the spin relaxation rate with $\delta\Gamma_{ij} = -\partial_k\beta_{ijk}$, which implies modifications of spin relaxation at surfaces and interfaces. Thus, the effect of the second term in Eq.~\eqref{S-SGE} can be absorbed in the redefinition of the spin relaxation rate in Eq.~\eqref{eq:S2}, which we assume in the following. 

To reveal the physics of the first term in Eq.~\eqref{S-SGE} and the significance of the pseudotensor $\gamma_{ij}$, we analyze the NLSM defined by the action $S = S_0 + S_2 + S_3^{(2)}$, where $S_0$ is the usual contribution given by the first three terms in Eq.~\eqref{eq:action0}. Let us concentrate on the saddle point of this action, $Q=g$. The saddle point condition yields the Usadel equation for the quasiclassical Green's function, 
\begin{align}
\label{eq:Usadel-SGE}
\partial_k\mathcal{J}_k+ [\tau_3(\hat{\omega} + i\mathbf{h}{\bm\sigma})+\hat{\Delta},g] = \mathcal{T} - \frac{1}{8}\Gamma_{jk}[\sigma_jg\sigma_k,g] \; ,   
\end{align}
where the matrix current $\mathcal{J}_k$ and matrix torque $\mathcal{T}$ are:
\begin{align}
    \label{eq:J-SGE}
    \mathcal{J}_k &= -Dg\tilde{\nabla}_kg + \frac{i}{16}\epsilon_{ijl}\gamma_{lk}
    \{[\sigma_i,g],\sigma_j + g\sigma_jg\}\,, \\
    \label{eq:T-SGE}
    \mathcal{T} &= i\alpha_{kj}[\sigma_{k},g\partial_{j}g] -\frac{i}{8}\epsilon_{ijl}\gamma_{lk} [\{\partial_kg,g\sigma_ig\},\sigma_j] \,.
\end{align}
By construction, these equations are consistent with the normalization condition $g^2=\mathbf{1}$ , where $\mathbf{1}$ is the identity matrix in Keldysh-Nambu-spin space. The retarded and advanced components of the general Usadel equation, Eq. (\ref{eq:Usadel-SGE}), reveal the spectral properties of the system, including spectral supercurrents. The Keldysh component provides a kinetic equation for distribution functions when the system is driven out of equilibrium. In the normal state, only the Keldysh component remains nontrivial \footnote{ In Supplemental Material Sec. \textcolor{purple}{IV} we derive from Eq.(\ref{eq:Usadel-SGE}) the charge and spin diffusion equations, which reveals explicitly  all possible magnetoelectric effects allowed by symmetry.}.  It can be expressed through the distribution functions or through the physical observables such as charge and spin accumulation, see Eqs. (S71-S78) in SM Sec. \ref{sec:KineticNormalState}.
 Importantly, anticommutators of $g$ with spin matrices in Eqs.~\eqref{eq:J-SGE}-\eqref{eq:T-SGE} induce coupling between singlet and triplet components of $g$. This indicates that the pseudotensor $\gamma_{ij}$ is related to magnetoelectric phenomena. The precise physics encoded in $\gamma_{ij}$ can be understood by analyzing Eqs.~\eqref{eq:J-SGE}-\eqref{eq:T-SGE} in the linear regime.

Let us assume that $g$ slightly deviates from a current-free 
and spin-independent value, $g=g_0+\delta g$, where $\{g_0,\delta g\}=0$ to ensure the normalization, $[g_0,\sigma_i]=0$ and $\delta g = \delta g_s + \delta g_t^i\sigma_i$. The matrix current and torque take the form,
\begin{align}
    \label{eq:J-linearSGE}
    \mathcal{J}_k &= -Dg_0\nabla_k\delta g + \gamma_{jk}\delta g^j_t \, , \\
    \label{eq:T-linearSGE}
    \mathcal{T} &= \gamma_{jk}\sigma_j\partial_k\delta g_s \, .
\end{align}
The second term in Eq.~\eqref{eq:J-linearSGE} describes a spin independent contribution to the matrix current proportional to the triplet (spin dependent) part of $\delta g$. Apparently, this corresponds to the spin-galvanic effect (SGE), also called the inverse Edelstein effect \cite{ganichev2019spin,levitov1985magnetoelectric,Aronov1989,Ivchenko1989,edelstein1990spin,Edelstein1995,yip2005magnetic,edelstein2005magnetoelectric,he2020magnetoelectric}.
Indeed, the physical current is obtained from the Keldysh component of the matrix current $j_k = \frac{\pi\nu}{2}{\rm Tr}\{\tau_3\mathcal{J}_k^K(t,t)\}$ \cite{bergeret2018colloquium}, which gives a linear  relation between  the charge current and the excess spin, which is defined as $\delta\mathbf{S} = \frac{\pi\nu}{2}{\rm Tr}\{\tau_3\mathbf{g}_t^{K}(t,t)\}$\cite{bergeret2018colloquium},
\begin{align}
    \label{eq:j-SGE}
    j_k^{\rm SGE} = \gamma_{ik}\delta S_i\; .
\end{align}
Therefore the pseudotensor $\gamma_{ij}$ in the action of the NLSM Eq.~\eqref{S-SGE} is the  spin-galvanic coefficient relating the charge current to the spin accumulation \cite{ganichev2019spin}. Interestingly, the SGE relation of Eq.~\eqref{eq:j-SGE} is highly universal -- it does not depend on the means used to create $\delta\mathbf{S}$, and has the same form in normal, superconducting, equilibrium, or nonequilibrium state. A similar universality has been noticed for models with linear SOC \cite{tokatly2017usadel}. Here we see that it is a universal property of SGE in all gyrotropic systems. In superconductors, the SGE manifests as $\phi_0$-effect and the superconducting diode effect \cite{nadeem2023superconducting,bobkova2022magnetoelectric}. Our theory thus provides a theoretical tool for the quantitative description of these effects in real materials, without relying on simplified models. 

The matrix torque in Eq.~\eqref{eq:T-linearSGE}, generated by the gradient of the singlet part of $\delta g$, is responsible for the inverse SGE also known as the current induced spin polarization or Edelstein effect. 
In the normal state, similarly to Rashba-Dresselhaus systems \cite{Gorini2017PRB}, it leads to the spin generation torque $T_i=\gamma_{ki}\partial_k\mu$, proportional to the gradient of electrochemical potential, in the spin diffusion equation. 

In superconductors, the matrix torque Eq.~\eqref{eq:T-linearSGE} acts as an effective Zeeman field proportional to a supercurrent, inducing a spin polarization of the condensate. As a nontrivial illustration of the superconducting inverse SGE and  the capability of our fully nonlinear Usadel equation in gyrotropic systems with SOC, we analyze the spin texture induced around a  vortex.

\begin{figure}
    \centering
    \includegraphics[width = 8.6cm]{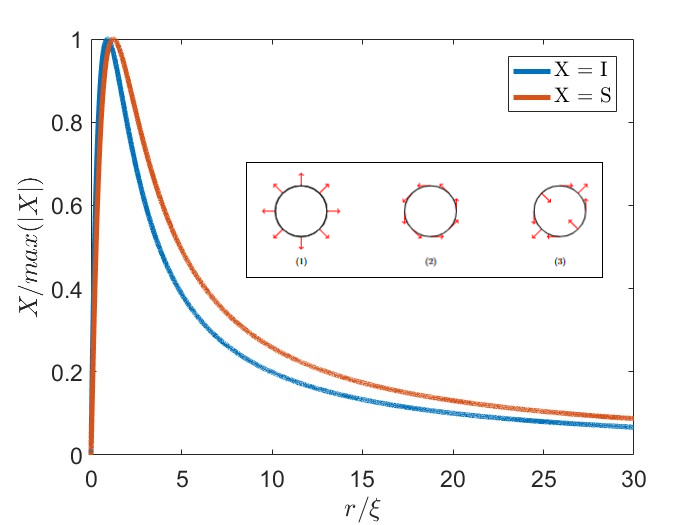}
    \caption{The current density ($I$) and spin density ($S$),  normalized to their maximal values, around a vortex with vorticity $n = 1$ as functions of the distance $r$,  in units of the coherence length $\xi$, from the vortex core.
    \textit{Inset:} The spin texture around the vortex for  several gyrotropic groups characterized by two nonzero elements of the spin-galvanic tensor, (1) $\gamma_{xy} = -\gamma_{yx}$ as in $C_{3v,4v,6v}$, (2) $\gamma_{xx} = \gamma_{yy}$ as  in $T,O,D_{4}$ and (3) $\gamma_{xx} = -\gamma_{yy}$ as  in $D_{2d},S_{4}$. 
    }
    \label{fig:smallTandlargeT}
\end{figure}

In a vortex, the phase of the pair amplitude only depends on the angle, while their magnitude depends on the radial coordinate $r$. As elaborated in SM Sec. \ref{sec:Vortex}, in Eq.~\eqref{eq:T-linearSGE} the phase derivative enters as an effective exchange field $\gamma_{kj}\hat{\phi}_{k}[\sigma_{j}\tau_{3}, g_{s}]$, where $\vec{\hat{\phi}}$ denotes the unit vector in the tangential direction, while the derivative with respect to the radial coordinate does not break time-reversal symmetry and hence does not contribute to the spin texture. Thus, the spin around the vortex points along $\gamma_{kj}\hat{\phi}_{k}$. 
The inset of Fig. \ref{fig:smallTandlargeT} sketches the 
textures for different point group symmetries.
The dependence of the current density and induced spin density as a function of the radial coordinate, has to be computed numerically by solving the Usadel equation, see SM Sec. \ref{sec:Vortex} for details.   Fig.~\ref{fig:smallTandlargeT} shows  the  result  for temperatures well below the critical temperature. 
The  relation between the two is nonlocal, since $S(r)$ is maximized further away from the core than the maximum current density. This nonlocality is reduced by increasing the temperature.

In the limit of high temperature,  $T\lesssim T_c$ the GL functional can be straightforwardly derived \cite{gor1959microscopic,golubov2003upper,ilic2022theory,houzet2015quasiclassical}.  In a gyrotropic system, the term $q_j\gamma_{ij}$ acts as an effective exchange field, which combines with $\mathbf{h}_i$ (where $q_k$ represents the  Fourier transform of the derivatives $i\partial_k$; see SM \ref{sec:GL}). Consequently, the GL free energy includes a term linear in derivatives known as the Lifshitz invariant, denoted as $F_L = \chi_{ij} h_j\gamma_{ik}\partial_k\varphi$, where $\chi_{ij}$ is the magnetic response function to a Zeeman field, inducing an excess spin, $\delta S_i=\chi_{ij}h_j$. This establishes a connection of the Lifshitz invariant to the universal spin-galvanic coefficient $\gamma_{ij}$ appearing in Eq.~\eqref{eq:j-SGE} and in the action Eq.~\eqref{S-SGE}. The full  list of all Lifshitz invariants for all gyrotropic point groups has been presented, for example, in Ref.~\cite{AgterbergInBook2012}.

In conclusion, we have presented a method for deriving the action of the Keldysh NLSM exclusively relying on the principles of charge conjugation, chronological and crystal symmetries. Applying this method, we have successfully derived the main objective of this paper, the quantum kinetic equation, commonly known  as the Usadel equation, for metallic systems featuring generic SOC. Through this derivation, we have identified the distinct terms and symmetries governing spin precession, relaxation, and spin-charge coupling.
 As a byproduct, the provided NLSM can be used to investigate superconducting fluctuations or localization effects in materials with arbitrary types of SOC by applying various techniques developed in the context of effective field theories to address properties of NLSM beyond the saddle point level. 
%Moreover, many techniques developed in the context of field theories, such as geometric considerations and renormalization group  analysis, can be applied to the effective field theory defined by the presented NLSM to explore the properties of materials with SOC beyond quasiclassical transport.
 
Our approach, while akin in spirit to the GL theory, offers an effective field theory valid  in the whole  temperature range. Moreover, it establishes a novel framework for deriving quantum kinetic theories applicable to diverse systems.

\section*{Acknowledgements}
We acknowledge discussions with Alexander Golubov, Pauli Virtanen, Stefan Ilic and  Tero Heikkil\"a.
We  acknowledge financial support from Spanish MCIN/AEI/
10.13039/501100011033 through projects PID2023-148225NB-C31, PID2023-148225NB-C32(SUNRISE),
and TED2021-130292B-C42, the Basque Government through grants IT-1591-22 and IT1453-22, and the 
European Union’s Horizon Europe research and innovation program under grant agreement No 101130224 (JOSEPHINE). 

%T.K. and F.S.B. acknowledge financial support from Spanish MCIN/AEI/
%10.13039/501100011033 through project PID2020-114252GB-I00 (SPIRIT)
%and TED2021-130292B-C42, and the Basque Government through grant IT-1591-22.
%I.V.T. acknowledges support by Grupos Consolidados UPV/EHU del Gobierno
%Vasco (Grant IT1453-22) and by the grant PID2020-112811GB-I00 funded
%by MCIN/AEI/10.13039/501100011033.

\bibliography{sources}
\appendix
\onecolumngrid\
\renewcommand{\theequation}{S\arabic{equation}}
\section{Charge conjugation symmetry}\label{sec:ChargeConjugation}
In the theoretical description of superconductivity, it is convenient to introduce  the bispinor of Fermion fields $\Psi = (\psi_{\uparrow},\psi_{\downarrow},\psi^{\dagger}_{\downarrow},-\psi^{\dagger}_{\uparrow})^{T}$ \cite{efetov1980interaction,altland2010condensed,kamenev2023field}. The bispinor $\Psi$ contains both $(\psi_{\uparrow},\psi_{\downarrow})$ and its time-reversed partner $(\psi_{\downarrow}^{\dagger},-\psi_{\uparrow}^{\dagger})$, where the time reversal operator is defined as $\sigma_{2}\mathcal{K}$. Complex conjugation is denoted by $\mathcal{K}$ and $\sigma_{2}$ is the second Pauli matrix in spin space. This is suitable for the description of superconductivity, which couples time-reversed pairs of electrons \cite{anderson1959theory}. The theory is most conveniently described in terms of $\Psi$ and $\Bar{\Psi} = \psi^{\dagger}\tau_{3}  = (\psi_{\uparrow}^{\dagger},\psi_{\downarrow}^{\dagger},-\psi_{\downarrow},\psi_{\uparrow})$ instead of $\Psi$ and $\Psi^{\dagger}$, so that any term that is the same for electrons and holes comes along with a $\tau_{0}$, while any term that is opposite for electrons and holes comes with $\tau_{3}$.

Consequently, not all degrees of freedom are independent. Indeed, $\bar{\Psi}$ and $\Psi$ are related by $\Psi = C\bar{\Psi}^{T}$, and hence $\bar{\Psi} = \Psi C^{-1}$, where $C = -i\tau_{1}\sigma_{2}$ is the so-called charge conjugation symmetry operator. Therefore, the matrix composed of $\Psi\bar{\Psi}$ necessarily satisfies the relation $C(\Psi(t')\bar{\Psi}(t))^{T}C^{-1} = C \bar{\Psi}^{T}(t) \Psi^{T}(t')C^{-1} = \Psi(t)\bar{\Psi}(t')$. Notably, this holds for all combinations of Fermion fields $\Psi,\bar{\Psi}$ in the domain of integration.

Since the Q-matrix is used for Hubbard-Stratonovich decoupling of $\Psi\Bar{\Psi}$ \cite{altland2010condensed,kamenev2009keldysh}, the domain of integration for the Q-matrix should be restricted to those Q-matrices that satisfy this same symmetry. In contrast to the main text, here we distinguish the Q-matrices on the contour from those written as matrices in Keldysh space by using the calligraphic $\mathcal{Q}$ on the contour and a noncalligraphic $Q$ in Keldysh space. On the contour the charge conjugation symmetry can be written in the same notation as the symmetry on $\Psi\Bar{\Psi}$:
\begin{align}
  \mathcal{Q}_{t_{1},t_{2}} = C\mathcal{Q}_{t_{2},t_{1}}^{T}C^{-1}\;,
  \end{align}
where $t_{1,2}$ are times on the contour \cite{gurarie2011single,ryu2010topological,schnyder2008classification}.

In the main text the nonlinear sigma model is, for convenience, presented in the rotated Keldysh space. First, the nonrotated Keldysh space is created by writing the different branches in matrix form, that is,
\begin{align}
Q'_{t,t'} = \begin{bmatrix}
    \mathcal{Q}_{t^{+},t'^{+}}&\mathcal{Q}_{t^{+},t'^{-}}\\\mathcal{Q}_{t^{-},t'^{+}}&\mathcal{Q}_{t^{-},t'^{-}}\;,
\end{bmatrix}
\end{align}
where the superscripts $\pm$ have been used to distinguish between times taken on the forward and backward branches of the contour and $t,t'$ are real times. The charge conjugation symmetry takes a form similar to that on the contour representation and may be written as 
\begin{align}
    Q'_{t,t '} = CQ_{t,t'}^{\prime T}C^{-1}\;.
\end{align}

The rotated Keldysh space can now be obtained via the transformation
\begin{align}
    Q_{t,t'} =  L\rho_{3}Q_{t,t'}'L^{\dagger}\;,
\end{align}
where $L = 1/\sqrt{2}(1-i\rho_{2})$ and $\rho_{i}$ is the $i$-th Pauli matrix in Keldysh space is performed because in this basis any causal function on the contour, such as the Green's function, is upper triangular in Keldysh space \cite{kamenev2009keldysh,belzig1999quasiclassical,levchenko2009transport,kamenev2023field,feigelman2000keldysh}. In this rotated Keldysh space the charge conjugation symmetry constraint on the manifold of integration can be written as

\begin{align}
    Q_{t,t'} = L\rho_{3}\Tilde{Q}_{t,t'}L^{-1} = L\rho_{3}C\Tilde{Q}_{t,t'}^{T}C^{-1}L^{-1} = L\rho_{3}CL^{-1}Q_{t,t'}^{T}L\rho_{3}C^{-1}L^{-1} = L\rho_{3}L^{-1}CQ_{t,t'}^{T}C^{-1}L\rho_{3}L^{-1}\;,
\end{align}
where the last equality follows from the fact that $C$ commutes with any Pauli matrix in $\rho$-space. We may now use $L\rho_{3}L^{-1} = \rho_{1}$ to find the form charge conjugation relation on $Q$ presented in the main text:
\begin{align}\label{eq:ArtificialDoublingSM}
    Q_{t,t'} =\rho_{1}CQ_{t,t'}^{T}C\rho_{1}  = \rho_{1}\tau_{1}\sigma_{2}Q_{t,t'}^{T}\sigma_{2}\tau_{1}\rho_{1}\;.
\end{align}
This symmetry is  referred to as the charge conjugation symmetry and introduced in Eq. (\textcolor{purple}{2}) in the main text. 
\newpage
\section{Chronological symmetry}\label{sec:chronological}
In this section we show that the condition $g = -\rho_{2}\tau_{3}g^{\dagger}\tau_{3}\rho_{2}$ and the consequent chronology symmetry of the action can be derived from the following two assumptions:

\begin{itemize}
     \item The contour ordered function $i\mathcal{G}_{C}= \langle \mathcal{T}_{C}\Psi^{\dagger}(t_{1})\Psi(t_{2})\rangle$, where $\langle\cdot\rangle$ is used to denote expectation values, is piecewise continuous as a function of $t_{1}$ or $t_{2}$ with its only discontinuity for $t_{1} = t_{2}$. 
     \item The contour ordered Green's function $i\mathcal{G}_{C} = \langle \mathcal{T}_{C}\Psi^{\dagger}(t_{1})\Psi(t_{2})\rangle$, where $\mathcal{T}_{C}$ denotes contour ordering on $C$ can be transformed into the anti-contour ordered function $i\mathcal{G}_{-C} = \langle \mathcal{T}_{-C}\Psi^{\dagger}(t_{1})\Psi(t_{2})\rangle$, where $\mathcal{T}_{-C}$ denotes anti-contour ordering on $C$ via Hermitian conjugation and adding a minus sign.
    
\end{itemize}
The latter condition is motivated by the generator of the time-evolution, the Hamiltonian, being Hermitian, in combination with the permutation rules for Grassmann variables \cite{lerner2003nonlinear,van2006introduction}. Continuity for $t_{1}\neq t_{2}$ is a consequence of the adiabatic assumption underlying the nonlinear sigma model formalism, while the possibility of a discontinuity at $t_{1} = t_{2}$ arises because of the difference in order of the operators in the expressions of $\mathcal{G}_{C}$ for $t_{1}<t_{2}$ and $t_{1}>t_{2}$.

 The first requirement, along with the definition of $i\mathcal{G}_{\pm C}$ can be written as \begin{align}&i\mathcal{G}_{C}(t_{1},t_{2}) = i\mathcal{G}^{>}(t_{1},t_{2})\theta_{C}(t_{1},t_{2})+i\mathcal{G}^{<}(t_{1},t_{2})\theta_{C}(t_{2},t_{1})\;,\nonumber\\&i\mathcal{G}_{-C}(t_{1},t_{2}) = i\mathcal{G}^{<}(t_{1},t_{2})\theta_{C}(t_{1},t_{2})+i\mathcal{G}^{>}(t_{1},t_{2})\theta_{C}(t_{2},t_{1})\;,\end{align} where $\theta_{C}$ is the contour Heaviside function and $i\mathcal{G}^{>}(t_{1},t_{2}) = \langle \Psi(t_{1})\Psi^{\dagger}(t_{2})\rangle$ and $i\mathcal{G}^{<}(t_{1},t_{2})= \langle \Psi^{\dagger}(t_{2})\Psi(t_{1})\rangle$ are continuous functions. 

The second requirement implies \begin{align}i\mathcal{G}_{-C} = -(i\mathcal{G}_{C})^{\dagger}\;,\\i\mathcal{G}^{>} = -(i\mathcal{G}^{<})^{\dagger}\;.\end{align}

Next to this we may define the time-ordered function $i\mathcal{G}^{\mathcal{T}}(t_{1},t_{2}) = \langle 
\mathcal{T}\psi(t_{1})\psi^{\dagger}(t_{2})\rangle$ where $\mathcal{T}$ denotes normal time-ordering of the real part of $t_{1,2}$, while $i\mathcal{G}^{\Tilde{\mathcal{T}}}(t_{1},t_{2}) = \langle \Tilde{\mathcal{T}}\psi(t_{1})\psi^{\dagger}(t_{2})\rangle$ is the anti-time-ordered version. We note that the conditions on $i\mathcal{G}_{\pm C}$ imply that \begin{align}i\mathcal{G}^{\Tilde{\mathcal{T}}} = -(i\mathcal{G}^{\mathcal{T}})^{\dagger}\;.\end{align} 

Next, we may write the contour ordered Green's function $\mathcal{G}_{C}$ in the nonrotated Keldysh space as, where, like in Sec. \ref{sec:ChargeConjugation} we distinguish in notation between contour representation and Keldysh space representation,
\begin{align}
    iG'(t,t') = \begin{bmatrix}
        i\mathcal{G}_{C}(t^{+},t^{+})&i\mathcal{G}_{C}(t^{+},t^{-})\\i\mathcal{G}_{C}(t^{-},t^{+})&i\mathcal{G}_{C}(t^{-},t^{-})
    \end{bmatrix} = \begin{bmatrix}
        i\mathcal{G}^{\mathcal{T}}(t,t')&i\mathcal{G}^{>}(t,t')\\i\mathcal{G}^{<}(t,t')&i\mathcal{G}^{\Tilde{\mathcal{T}}}(t,t')
    \end{bmatrix}\;.
\end{align}
Thus, the above restrictions on $i\mathcal{G}_{C}$ may be written in terms of $iG'$ as 
\begin{align}
iG' = -\rho_{1}(iG)^{\prime\dagger}\rho_{1}\;.
\end{align}

Next we apply the Keldysh rotation $iG = L\rho_{3}iG'L^{-1}$. Using the rules for products of Pauli matrices, along with $L^{-1} = L^{\dagger}$, we find
\begin{align}
iG&=  -L\rho_{3}\rho_{1}(iG)^{\prime\dagger}\rho_{1}L^{-1}=-L\rho_{2}(iG)^{\prime\dagger}\rho_{3}\rho_{2}L^{-1} = -\rho_{2}L(iG)^{\prime\dagger}\rho_{3}L^{-1}\nonumber\\& = -\rho_{2}(L\rho_{3}(iG)^{\prime}L^{-1})^{\dagger}\rho_{2} = -\rho_{2}(iG)^{\dagger}\rho_{2}\;.
\end{align}

The quasiclassical Green's function is defined using $\bar{\Psi} = \Psi^{\dagger}\tau_{3}$, so that any single-particle term that acts the same on an electrons and its time-reversed partner comes with $\tau_{0}$, while any term that acts differently on them comes with $\tau_{3}$. Therefore, the quasiclassical Green's function $g = \langle\Psi\bar{\Psi}\rangle$ can be written as $g = iG\tau_{3}$. Consequently, the chronological symmetry requirement on $g$ takes the form
\begin{align}
    g = -\tau_{3}\rho_{2}g^{\dagger}\rho_{2}\tau_{3}\;.
\end{align}
Hence, in terms of Q-matrices we find
\begin{align}
    \int[dQ] Q e^{iS[Q]} = \int[dQ] -\tau_{3}\rho_{2}Q^{\dagger}\rho_{2}\tau_{3} (e^{iS[Q]})^{*} = \int d\Tilde{Q} \Tilde{Q} e^{(iS[-\tau_{3}\rho_{2}\Tilde{Q}^{\dagger}\rho_{2}\tau_{3}])^{*}}\;.
\end{align}
Replacing $\Tilde{Q}$ by $Q$ in the last expression we conclude that this we may impose the symmetry $iS[Q] = (iS[-\tau_{3}\rho_{2}Q^{\dagger}\rho_{2}\tau_{3}])^{*}$ on the action ({\it cf.} Eq. (\textcolor{purple}{3}) in the main text). 
\newpage

\section{Terms combining Pauli matrices and derivatives}\label{sec:Mainterms}
In this section, we derive the different terms leading to Eqs. (\textcolor{purple}{4}-\textcolor{purple}{6}) in the main text. For this, we determine the allowed terms in the action to second and third order in $\sigma$-matrices and derivatives.
\subsection{Second order terms}
 Terms containing two Pauli matrices or two derivatives, provide the general relaxation, $\Gamma_{ij}$, and diffusion tensors, $D_{ij}$, in Eq. (\textcolor{purple}{4}) in the main text. Here we focus on the derivation of other terms that contain one Pauli matrix and one derivative.
 There are two of such  terms,
 \begin{align}\bar{\alpha}_{kj}\sigma_{k}\partial_{j}Q\;,\end{align} and \begin{align}\frac{i}{2}\alpha_{kj}\sigma_{k}Q\partial_{j}Q\;.\end{align} Applying transposition to the first we find
\begin{align}
    \text{Tr}-\partial_{j}Q^{T}\sigma_{y}\sigma_{k}\sigma_{y} =-\text{Tr}\rho_{1}\tau_{1}\sigma_{y}\partial_{j}Q^{T}\sigma_{y}\rho_{1}\tau_{1}\sigma_{k} = -\text{Tr}\partial_{j}Q\sigma_{k} = -\text{Tr}\sigma_{k}\partial_{j}Q\;.
\end{align}
To obtain a charge conjugation symmetric term, this term should be equivalent to the one we started from, hence this imposes the restriction $\bar{\alpha}_{kj} = -\bar{\alpha}_{kj}$ on the corresponding tensor, i.e. $\bar{\alpha}_{kj} = 0$. We conclude that this term is not allowed in the action.

Next we apply charge conjugation on the second term. It gives
\begin{align}
    \text{Tr}\partial_{j}Q^{T}Q^{T}\sigma_{y}\sigma_{k}\sigma_{y} =-\text{Tr}\rho_{1}\tau_{1}\sigma_{y}\partial_{j}Q^{T}\sigma_{y}\rho_{1}\tau_{1}\sigma_{y}\rho_{1}\tau_{1}Q^{T}\tau_{1}\rho_{1}\sigma_{y}\sigma_{k} = -\text{Tr}\partial_{j}QQ\sigma_{k} = \text{Tr}\sigma_{k}Q\partial_{j}Q\;.
\end{align}
Thus, this term is charge conjugation symmetric for any second rank pseudotensor $\Tilde{\alpha}_{kj}$.

Next, we consider the influence of chronology symmetry. We find
\begin{align}
     \text{Tr}\partial_{j}Q^{\dagger}Q^{\dagger}\sigma_{k}  = \text{Tr}\tau_{3}\rho_{2}\partial_{j}Q^{\dagger}\rho_{2}\tau_{3}\tau_{3}\rho_{2}Q^{\dagger}\rho_{2}\tau_{3}\sigma_{k} = \text{Tr}\partial_{j}\Tilde{Q}\Tilde{Q}\sigma_{k} = -\text{Tr}\sigma_{k}\Tilde{Q}\partial_{j}\Tilde{Q}\;,
\end{align}
where $\Tilde{Q} = -\rho_{2}\tau_{3}Q^{\dagger}\tau_{3}\rho_{2}$. Since $iS(Q)^{*}= iS(\Tilde{Q})$ it now follows that $\alpha_{kj} = \alpha_{kj}^{*}$, that is, $\alpha_{kj}$ is a real second rank pseudotensor. The corresponding term in the bulk equation, which can be derived via the requirement that at the saddle point the action is invariant under variations $g+[\delta g,g]$:
\begin{align}
    i\text{Tr}\Big(\alpha_{kj}\sigma_{k}[\delta g,g]\partial_{j}g+\alpha_{kj}\sigma_{k}g\partial_{j}[\delta g,g]\Big)\;.
\end{align}
If we use the cyclic property of the trace to write $\delta g$ and the end, we obtain
\begin{align}
    -\text{Tr}\Bigg(i\alpha_{kj}\Big(\sigma_{k}\partial_{j}g\sigma_{k}g-g\partial_{j}g\sigma_{k}+\sigma_{k}g\partial_{j}g-\partial_{j}g\sigma_{k}g\Big)\delta g+i\alpha_{kj}\Big(\sigma_{k}-g\sigma_{k}g\Big)\partial_{j}\delta g\Bigg)\;.
\end{align}
After doing integration by parts, we obtain the bulk contribution
\begin{align}
    -\text{Tr}\Big(i\alpha_{kj}[\sigma_{k},g\partial_{j}g]+\partial_{j}(i\alpha_{kj}g[\sigma_{k},g])\Big)\delta g\;.
\end{align}
%\begin{align}
%    i\alpha_{kj}[[\sigma_{k},\partial_{j}g],g].
%\end{align}

%Because the action contains terms with derivatives, integration by parts is necessary to derive the saddle point condition. This additionally yields terms that are evaluated only at the boundary. These terms have the form $n_{j}J_{j}$, where $n_{j}$ is the normal vector of the boundary and $J_{j}$ is the contribution to the current of the term under consideration. Equivalently, the expression for the current can be found by variation of the action with the vector potential. 
It is convenient to separate this contribution into the current and the torque. The current may be obtained via variation of the action with the vector potential or equivalently by evaluation of the boundary terms that arise upon the required integration by parts in the derivation of the saddle point.
The contribution of the term $i\alpha_{kj}\sigma_{k}Q\partial_{j}Q$ to the current is
\begin{align}
    J_{j} = i\alpha_{kj}g[\sigma_{k},g]\;.
\end{align}
Consequently, the torque reads
\begin{align}
    \mathcal{T} = -i\alpha_{kj}[\sigma_{k},g\partial_{j}g]\;.
\end{align}
%In the normal state the Keldysh component of this expression vanishes.
\subsection{Third order terms with  two Pauli matrices and  one derivative}
There exist two terms of this order: \begin{align}\beta_{ijk}\sigma_{i}Q\sigma_{j}\partial_{k}Q\end{align} and \begin{align}\textcolor{blue}{-}\frac{i}{8}\gamma_{ijk}\sigma_{i}Q\sigma_{j}Q\partial_{k}Q\end{align}.
Transposition of the first term gives
\begin{align}
    \text{Tr}\partial_{k}Q^{T}\sigma_{y}\sigma
    _{j}\sigma_{y}Q^{T}\sigma_{y}\sigma_{i}\sigma_{y} = \text{Tr}\rho_{1}\tau_{1}\sigma_{y}\partial_{k}Q^{T}\sigma_{y}\rho_{1}\tau_{1}\sigma
    _{j}\tau_{1}\rho_{1}\sigma_{y}Q^{T}\sigma_{y}\tau_{1}\rho_{1}\sigma_{i}  = \text{Tr}\partial_{k}Q\sigma_{j}Q\sigma_{i} = \text{Tr}\sigma_{j}Q\sigma_{i}\partial_{k}Q\;,
\end{align}
that is, charge conjugation symmetry requires that the corresponding tensor $\beta_{ijk}$ satisfies $\beta_{ijk} = \beta_{jik}$.

To find the restriction by chronological symmetry we calculate
\begin{align}
    \text{Tr}\partial_{k}Q^{\dagger}\sigma_{j}Q^{\dagger}\sigma_{i} = \text{Tr}\rho_{2}\tau_{3}\partial_{k}Q^{\dagger}\rho_{2}\tau_{3}\sigma_{j}\rho_{2}\tau_{3}Q^{\dagger}\rho_{2}\tau_{3}\sigma_{i} = \text{Tr}\partial_{k}\Tilde{Q}\sigma_{j}\Tilde{Q}\sigma_{i} = \text{Tr}\sigma_{j}\Tilde{Q}\sigma_{i}\partial_{k}Q\;,
\end{align}
that is $\beta_{ijk} = \beta_{jik}^{*} = \beta_{ijk}$, where in the last equality we used the charge conjugation symmetry relation derived above.  Thus, this tensor is real. The contribution to the bulk equation can be found using variation:
\begin{align}
    &\beta_{ijk}\sigma_{i}[\delta g,g]\sigma_{j}\partial_{k}g+\sigma_{i}g\sigma_{j}\partial_{k}[\delta g,g]\\
    &=\beta_{ijk}\Big(g\sigma_{j}\partial_{k}g\sigma_{i}-\sigma_{j}\partial_{k}g\sigma_{i}g+\partial_{k}g\sigma_{i}g\sigma_{j}-\sigma_{i}g\sigma_{j}\partial_{k}g\Big)\delta g + \Big(g\sigma_{i}g\sigma_{j}-\sigma_{i}g\sigma_{j}g\Big)\partial_{k}g\\
    &=\Bigg(-\beta_{ijk}\partial_{k}[\sigma_{i}g\sigma_{j},g]+\partial_{k}(\beta_{ijk}[\sigma_{i}g\sigma_{j},g])\Bigg)\delta g\;.
\end{align}
Specifically, if $\beta_{ijk}$ is constant, this reduces to
\begin{align}
     0\;.
\end{align}
by symmetry in the $i,j$ indices. Indeed, in that case we may write this term as a total derivative term $\partial_{k}(\sigma_{i}Q\sigma_{j}Q)$, and hence this term can only enter via the boundaries.

Its contribution to the current is
\begin{align}
    J_{k} = -\beta_{ijk}[\sigma_{j}g\sigma_{i},g]\;.
\end{align}
%In the normal state the Keldysh component is 
%\begin{align}
%J_{k}^{K} = 2\beta_{ijk}(\{F,\sigma_{j}\sigma_{i}\}-2\sigma_{j}F\sigma_{i}).
%\end{align}
%By symmetry of $\beta_{ijk}$ upon exchange of $i,j$ we may write this as
%\begin{align}
%    J_{k}^{K} = 4\beta_{ijk}(F-\sigma_{j}F\sigma_{i})
%\end{align}
Consequently, the torque is given by
\begin{align}
    \mathcal{T} = -\beta_{ijk}\partial_{k}[\sigma_{j}g\sigma_{i},g]\;.
\end{align}

The second term is $\sigma_{i}Q\sigma_{j}Q\partial_{k}Q$. We have, using charge conjugation symmetry:
\begin{align}
    &\text{Tr}\partial_{k}Q^{T}Q^{T}\sigma_{y}\sigma
    _{j}\sigma_{y}Q^{T}\sigma_{y}\sigma_{i}\sigma_{y} = \text{Tr}\rho_{1}\tau_{1}\sigma_{y}\partial_{k}Q^{T}\sigma_{y}\rho_{1}\tau_{1}\sigma_{y}\rho_{1}\tau_{1}Q^{T}\tau_{1}\rho_{1}\sigma_{y}\sigma
    _{j}\tau_{1}\rho_{1}\sigma_{y}Q^{T}\sigma_{y}\tau_{1}\rho_{1}\sigma_{i}\nonumber\\ & = \text{Tr}\partial_{k}Q Q\sigma_{j}Q\sigma_{i} = -\text{Tr}\sigma_{j}Q\sigma_{i}Q\partial_{k}Q\;,
\end{align}
that is, the third rank pseudotensor corresponding to this term is antisymmetric in the first two indices and we may write it as $\frac{1}{8}\epsilon_{ijl}\gamma_{lk}$, where $\epsilon_{ijl}$ is the fully antisymmetric third rank pseudotensor and $\Tilde{\gamma}_{lk}$ is a second rank pseudotensor. The $\frac{1}{8}$ is chosen for convenience.

For chronological symmetry we take the Hermitian conjugate and calculate
\begin{align}
    \text{Tr}\partial_{k}Q^{\dagger}Q^{\dagger}\sigma_{j}Q^{\dagger}\sigma_{i} = \text{Tr}\tau_{3}\rho_{2}\partial_{k}Q^{\dagger}\tau_{3}\rho_{2}Q^{\dagger}\tau_{3}\rho_{2}\sigma_{j}\tau_{3}\rho_{2}Q^{\dagger}\tau_{3}\rho_{2}\sigma_{i} = -\text{Tr}\partial_{k}\Tilde{Q}\Tilde{Q}\sigma_{j}\Tilde{Q}\sigma_{i} = -\text{Tr}\sigma_{j}\Tilde{Q}\sigma
    _{i}\Tilde{Q}\partial_{k}\Tilde{Q}\;,
\end{align}
that is, $\gamma_{lk}$ is a real second pseudotensor.

The contribution to the bulk Usadel equation can be calculated using variation again
\begin{align}
    &-\frac{i}{8}\epsilon_{ijl}\gamma_{lk}\Big(\sigma_{i}[\delta g,g]\sigma_{j}g\partial_{k}g+\sigma_{i}g\sigma_{j}[\delta g,g]\partial_{k}g+\sigma_{i}g\sigma_{j}g\partial_{k}[\delta g,g]\Big)\\
    &=-\frac{i}{8}\epsilon_{ijl}\gamma_{lk}\Big(g\sigma_{j}g\partial_{k}g\sigma_{i}-\sigma_{j}g\partial_{k}g\sigma_{i}g+g\partial_{k}g\sigma_{i}g\sigma_{j}-\partial_{k}g\sigma_{i}g\sigma_{j}g+\partial_{k}g\sigma_{i}g\sigma_{j}g-\sigma_{i}g\sigma_{j}g\partial_{k}g\Big)\delta g\nonumber\\&-\frac{i}{8}\epsilon_{ijl}\gamma_{lk}\Big(g\sigma_{i}g\sigma_{j}g-\sigma_{i}g\sigma_{j}\Big)\partial_{k}g\;.
\end{align}
In the first bracket, the fourth and fifth term cancel. The first and sixth term can be written as a commutator with $\sigma_{i}$, while the second and third can be written as a commutator with $\sigma_{j}$.
After integration by parts, and using that each term changes sign upon interchanging $i,j$, we obtain
\begin{align}
    -\frac{i}{8}\epsilon_{ijl}\gamma_{lk}\Big([g\partial_{k}g\sigma_{i}g-g\sigma_{i}g\partial_{k}g,\sigma_{j}]\Big)-\partial_{k}(\frac{i}{8}\epsilon_{ijl}\gamma_{lk}[\sigma_{i}g\sigma_{j}g,g])\;.
\end{align}
%For constant $\gamma_{lk}$ this reduces to
%\begin{align}
%    \frac{i}{8}\epsilon_{ijl}\gamma_{lk}[\sigma_{j}g\partial_{k}g\sigma_{i}+\partial_{k}g\sigma_{i}g\sigma_{j}-\sigma_{j}\partial_{k}g\sigma_{i}g-\sigma_{i}g\sigma_{j}\partial_{k}g,g] .
%\end{align}
%In the normal state this becomes
%\begin{align}
%    2\gamma_{lk}\tau_{3}\{\partial_{k}F,\sigma_{c}\}.
%\end{align}
%\textcolor{blue}{There is a minus sign discrepancy here?}
The contribution to the current is
\begin{align}
    J_{k} = \frac{i}{8}\epsilon_{ijl}\gamma_{lk}[\sigma_{i}g\sigma_{j}g,g]= \frac{i}{16}\epsilon_{ijl}\gamma_{lk}\{\sigma_{j}+g\sigma_{j}g,[\sigma_{i},g]\}\;.
\end{align}
and the torque of this term is
\begin{align}
    \mathcal{T} = -\frac{i}{8}\epsilon_{ijl}\gamma_{lk}[\{\partial_{k}g,g\sigma_{i}g\},\sigma_{j}]\;.
\end{align}
{These two last equations correspond to the contribution proportional to the pseudotensor $\gamma_{ij}$ in  the expressions   Eqs. (8-9) of the main text.}
%In the normal state the Keldysh component of this expression becomes
%\begin{align}
%    J_{k}^{K} = i\epsilon_{ijl}\tau_{3}\gamma_{lk}(2\sigma_{i}F\sigma_{j}-\{\sigma_{i}\sigma_{j},F\}).
%\end{align}
%Thus,
%\begin{align}
%    \text{Tr}J_{k}^{K} = %4\gamma_{lk}\text{Tr}\tau_{3}\sigma_{l}F,
%\end{align}
\subsection{Third order terms with one Pauli matrix and  two derivatives}
There exists two such terms; one with two $Q$'s:\begin{align}-i\frac{D}{4}\varkappa_{ijk}\sigma_{i}\partial_{j}Q\partial_{k}Q\;,\end{align} and one with three $Q$'s: \begin{align}i\frac{D}{4}\theta_{ijk}\sigma_{i}Q\partial_{j}Q\partial_{k}Q\;.\end{align}
The factors $\frac{D}{4}$ are chosen so that the $\chi_{al}$ corresponds to the spin-Hall and spin-swapping coefficients in the normal state, as shown in section \ref{sec:RelationUsualDef}.
Charge conjugation on the first term implies
\begin{align}
    -\partial_{k}Q^{T}\partial_{j}Q^{T}\sigma_{y}\sigma_{i}\sigma_{y} =-\sigma_{y} \rho_{1}\tau_{1}\partial_{k}Q^{T}\tau_{1}\rho_{1}\rho_{1}\tau_{1}\partial_{j}Q^{T}\tau_{1}\rho_{1}\sigma_{y}\sigma_{i}  = -\text{Tr}\partial_{k}Q\partial_{j}Q\sigma_{i} = -\text{Tr}\sigma_{i}\partial_{k}Q\partial_{j}Q\;,
\end{align}
that is, $\varkappa_{ijk} = -\varkappa_{ikj}$ and hence we may write it as $\epsilon_{jkl}\varkappa_{il}$, where $\epsilon_{jkl}$ is the third rank antisymmetric pseudotensor and $\varkappa_{il}$ is a second rank tensor. 
chronological symmetry gives
\begin{align}
    \partial_{k}Q^{\dagger}\partial_{j}Q^{\dagger}\sigma_{i} = \tau_{3}\rho_{2}\partial_{k}Q^{\dagger}\rho_{2}\tau_{3}\tau_{3}\rho_{2}\partial_{j}Q^{\dagger}\rho_{2}\tau_{3}\sigma_{i} = \partial_{k}\Tilde{Q}\partial_{j}\Tilde{Q}\sigma_{i} = \sigma_{i}\partial_{k}\Tilde{Q}\partial_{j}\Tilde{Q}\;,
\end{align}
that is, $\varkappa_{il}$ needs to be a real second rank tensor.

The contribution to the bulk equation can be found using variation:
\begin{align}
    &=-i\frac{D}{4}\epsilon_{jkl}\varkappa_{il}\text{Tr}\Bigg(\sigma_{i}\partial_{j}[\delta g,g]\partial_{k}g+\sigma_{i}\partial_{j}g\partial_{k}[\delta g,g]\Bigg)\\
    &=i\frac{D}{4}\epsilon_{jkl}\varkappa_{il}\text{Tr}\Bigg(\sigma_{i}\partial_{k}[\delta g,g]\partial_{j}g-\sigma_{i}\partial_{j}g\partial_{k}[\delta g,g]\Bigg)\\
    &=i\frac{D}{4}\epsilon_{jkl}\varkappa_{il}\text{Tr}\Bigg(\Big(\partial_{k}g\partial_{j}g\sigma_{i}-\partial_{j}g\sigma_{i}\partial_{k}g-\partial_{k}g\sigma_{i}\partial_{j}g+\sigma_{i}\partial_{j}g\partial_{k}g\Big)\delta g + \Big(g\partial_{j}g\sigma_{i}-\partial_{j}g\sigma_{i}g-g\sigma_{i}\partial_{j}g+\sigma_{i}\partial_{j}g g\Big)\partial_{k}\delta g\Bigg)\\
    &=\text{Tr}\Bigg(\Big(i\frac{D}{4}\epsilon_{jkl}\varkappa_{il}[\sigma_{i},\partial_{j}g\partial_{k}g]-\partial_{k}(i\frac{D}{4}\epsilon_{jkl}\varkappa_{il}[[\sigma_{i},\partial_{j}g],g])\Big)\delta g\Bigg)\;.
\end{align}
For constant $\varkappa_{il}$ this reduces to
\begin{align}
     0\;.
\end{align}
due to the antisymmetry in the $j,k$ indices. 
The contribution to the current is
\begin{align}
    J_{k} =i\frac{D}{4}\varkappa_{il}(\epsilon_{jki}[\sigma_{i}\partial_{j}g,g]+\epsilon_{kji}[\partial_{j}g\sigma_{i},g]) =i\frac{D}{4}  \varkappa_{il}\epsilon_{jkl}[[\sigma_{i},\partial_{j}g],g]\;.
\end{align}
Note that we need to swap the $j,k$ indices for the second term to get a current contribution to $J_{k}$ and not $J_{j}$.
%In the normal state this expression becomes
%\begin{align}
%    J_{k}^{K} = -iD\varkappa_{il}[\sigma_{i},\partial_{j}F]
%\end{align}
Because the contribution to the bulk equation vanishes the torque satisfies
\begin{align}
    \mathcal{T} = i\frac{D}{4}\epsilon_{jkl}\varkappa_{il}[\sigma_{i},\partial_{j}g\partial_{k}g]  = i\frac{D}{4}\varkappa_{il}\varepsilon_{jki}\partial_{k}[[\sigma_{i},\partial_{j}g],g]\;.
\end{align}

For the second term we obtain
\begin{align}
   & -\text{Tr}\partial_{k}Q^{T}\partial_{j}Q^{T}Q^{T}\sigma_{y}\sigma_{i}\sigma_{y} =-\text{Tr}\sigma_{y} \rho_{1}\tau_{1}\partial_{k}Q^{T}\tau_{1}\rho_{1}\sigma_{y}\sigma_{y}\rho_{1}\tau_{1}\partial_{j}Q^{T}\tau_{1}\rho_{1}\sigma_{y}\sigma_{y}\rho_{1}\tau_{1}Q^{T}\tau_{1}\rho_{1}\sigma_{y}\sigma_{i}  \nonumber\\&= -\text{Tr}\text{Tr}\partial_{k}Q\partial_{j}QQ\sigma_{i} = -\text{Tr}\sigma_{i}Q\partial_{k}Q\partial_{j}Q\;.
\end{align}
Thus, the pseudotensor of this term satisfies $\theta_{ijk} = -\theta_{ikj}$ and we may write the tensor belonging to this term as $\epsilon_{jkl}\theta_{il}$, where $\theta_{il}$ is a second rank tensor. The choice for the factor $\frac{D}{4}$ was made to ensure that that in an isotropic material our tensor $\theta_{il}$ reduces to the conventional spin-Hall angle, as shown in section \ref{sec:RelationUsualDef}. 

Chronological symmetry gives
\begin{align}
    \text{Tr}\partial_{k}Q^{\dagger}\partial_{j}Q^{\dagger}Q^{\dagger}\sigma_{i} = \text{Tr}\rho_{2}\tau_{3}\partial_{k}Q^{\dagger}\tau_{3}\rho_{2}\rho_{2}\tau_{3}\partial_{j}Q^{\dagger}\tau_{3}\rho_{2}\rho_{2}\tau_{3}Q^{\dagger}\tau_{3}\rho_{2}\sigma_{i}  = -\text{Tr}\partial_{k}\Tilde{Q}\partial_{j}\Tilde{Q}\Tilde{Q}\sigma_{i} = \text{Tr}\sigma_{i} \partial_{k}\Tilde{Q}\partial_{j}\Tilde{Q}\Tilde{Q} =\text{Tr}\sigma_{i} \Tilde{Q}\partial_{k}\Tilde{Q}\partial_{j}\Tilde{Q}  \;,
\end{align}
and hence $\theta_{il}$ is a real second rank tensor. 
Its contribution to the bulk equation can be found using
\begin{align}
    &\frac{D}{4}\epsilon_{jkl}\theta_{il}\sigma_{i}\Big([\delta g,g]\partial_{j}g\partial_{k}g+g\partial_{j}[\delta g,g]\partial_{k}g+g\partial_{j}g\partial_{k}[\delta g,g]\Big)\\
    &\frac{D}{4}\epsilon_{jkl}\theta_{il}\sigma_{i}\Big([\delta g,g]\partial_{j}g\partial_{k}g-g\partial_{k}[\delta g,g]\partial_{j}g+g\partial_{j}g\partial_{k}[\delta g,g]\Big)\\
    &=\frac{D}{4}\epsilon_{jkl}\theta_{il}\Bigg(\Big(g\partial_{j}g\partial_{k}g\sigma_{i}-\partial_{j}g\partial_{k}g\sigma_{i}g-\partial_{k}g\partial_{j}g\sigma_{i}g+\partial_{j}g\sigma_{i}g\partial_{k}g+\partial_{k}g\sigma_{i}g\partial_{j}g-\sigma_{i}g\partial_{j}g\partial_{k}g\Big)\delta g\nonumber\\&+\Big(-g\partial_{j}g\sigma_{i}g+\partial_{j}g\sigma_{i}+g\sigma_{i}g\partial_{j}g+\sigma_{i}\partial_{j}g\Big)\partial_{k}g\Bigg)\\
    &= -\Bigg(\frac{D}{4}\epsilon_{jkl}\theta_{il}\Big([\sigma_{i},g\partial_{j}g\partial_{k}g]\Big)+\partial_{k}\Big(\frac{D}{4}\epsilon_{jkl}\theta_{il}\{\sigma_{i}+g\sigma_{i}g,\partial_{j}g\}\Big)\Bigg)\;,
\end{align}
where it was used that in the first bracket the second and the third term cancel, just like the fourth and the fifth, due to the antisymmetry in $j,k$.
%\begin{align}
%    -\frac{D}{4}\epsilon_{jkl}\theta_{il}[\partial_{j}g\partial_{k}g\sigma_{i}-\partial_{k}g\sigma_{i}\partial_{j}g-\sigma_{i}\partial_{k}g\partial_{j}g,g],
%\end{align}

%where it was used that the $\partial_{j}\partial_{k}g$-terms drop out due to antisymmetry in $j,k$.
In the normal state this term drops out of the Usadel equation, since only one of the derivative terms can act on a Keldysh component while the retarded and advanced components are constant in the normal state. The contribution to the current is
\begin{align}
    J_{k} = -\frac{D}{4}\theta_{il}(\epsilon_{jkl}[\sigma_{i}g\partial_{j}g,g]+\epsilon_{kjl}[\partial_{j}g\sigma_{i}g,g]) = -\frac{D}{4}\epsilon_{jkl}\theta_{il}[[\sigma_{i}g,\partial_{j}g],g] = \frac{D}{4}\epsilon_{jkl}\theta_{il}\{\sigma_{i}+g\sigma_{i}g,\partial_{j}g\}\;.
\end{align}
%In the normal state, where $g^{K} = 2\tau_{3}F$ this becomes
%\begin{align}
%    J_{k}^{K} = \epsilon_{jkl}D\theta_{il}\tau_{3}\{\sigma_{i},\partial_{j}F\}.
%\end{align}
Thus, the torque is given by
\begin{align}
    \mathcal{T} = -\frac{D}{4}\varepsilon_{jkl}\theta_{il}[\sigma_{i},g\partial_{j}g\partial_{k}g]\;.
\end{align}
\subsubsection{ Isotropic materials }\label{sec:RelationUsualDef}
The above contributions to the matrix current, C22, C23, C27,C28, have been derived in the most general case. If the system is isotropic, 
 the only allowed second rank tensor is the identity tensor, and hence we may write and $\theta_{il} = \theta\delta_{il}$ and $\varkappa_{il} = \varkappa\delta_{il}$. In this case the current in the normal state may be written as
\begin{align}
    J_{k}^{K} = -2D\text{Tr}\partial_{k}F +D\theta\epsilon_{jki}\tau_{3}\{\sigma_{i},\partial_{j}F\}-iD\varkappa\epsilon_{jki}[\sigma_{i},\partial_{j}F]\;.
\end{align}
Next, we define $j_{k} = \text{Tr}\tau_{3}J_{k}^{K}$ and $j_{ki} = \text{Tr}\sigma_{i}J_{k}^{K}$. Moreover, we define $j_{k}^{(0)} = -D\text{Tr}(g\nabla g)^{K} = -2D\text{Tr}\partial_{k}F$ and $j_{ki}^{(0)} = -2D\text{Tr}\sigma_{i}\partial_{k}F$. With this we may write
\begin{align}
    \text{Tr}\sigma_{l}D\theta\epsilon_{jki}\tau_{3}\{\sigma_{i},\partial_{j}F\} = 2D\theta\epsilon_{jki}\delta_{il}\text{Tr}\tau_{3}\sigma_{l}\partial_{j}f  = -\theta\epsilon_{jki }j_{ji}^{(0)}\;,
\end{align}
while for the spin-swapping term we may write
\begin{align}
    -iD\varkappa\epsilon_{jki}\text{Tr}\sigma_{l}[\sigma_{i},\partial_{j}F] = -2iD\varkappa\epsilon_{jki} i\epsilon_{lim}\text{Tr}\sigma_{m}\partial_{j}F = -2D\varkappa\epsilon_{jki}\epsilon_{ilm}\text{Tr}\sigma_{m}\partial_{j}F = \varkappa\epsilon_{jki}\epsilon_{ilm}j_{jm}^{(0)}\;,
\end{align}
where in the second to last equality we used $i^{2} = -1$ and $\epsilon_{lim} = -\epsilon_{ilm}$.
For $k = l$ we have $\epsilon_{jli}\epsilon_{ilm}j_{jm}^{(0)} = -\epsilon_{jli}\epsilon_{mli}j_{jm}^{(0)} = \sum_{m\neq l}j_{mm}^{(0)}$. On the other hand, for $k\neq l$ we have $\epsilon_{jki}\epsilon_{ilm}j_{jm}^{(0)} = \epsilon_{jki}\epsilon_{lmi}j_{jm}^{(0)} = j_{mk}^{(0)}$. Hence we may write the contribution of the spin-swapping term as $j_{mk}^{(0)}-\delta_{kl}j_{mm}^{(0)}$.
With this we may express the current as
\begin{align}
    j_{kl} = \text{Tr}\sigma_{l}J_{k}^{K} = j_{kl}^{(0)}-\theta\epsilon_{jki }j_{ji}^{(0)}+\varkappa(j_{lk}^{(0)}-\delta_{kl}j_{mm}^{(0)})\;,
\end{align}
that is, $\theta$ and $\varkappa$ are the spin-Hall and spin-swapping coefficients in isotropic systems consistent with previous works \cite{khaetskii2008spin,lifshits2009swapping}.
\newpage
\section{Keldysh structure and kinetic equation}\label{sec:KineticNormalState}
In this section we give a brief outline of the Keldysh-structure of the Green's functions used in this letter, based on \cite{kamenev2009keldysh}, and present the kinetic equations in the normal state.

Within the Keldysh contour formalism, one uses a contour that runs along the time-axis between $\pm\infty$ in both directions. This ensures that the denominator of the generating function of the system becomes trivial by virtue of cancelling contributions from the two oppositely directed branches. Moreover, because the contour runs along the real-time axis, the formalism gives time-dependent equations and is capable of describing nonequilibrium physics.

It is convenient to write these two different branches in a matrix structure, that is, to define the Green's function on the contour as
\begin{align}\label{eq:ContourGreensfunction}
    \Tilde{g}(t,t') = \begin{bmatrix}
        \Tilde{g}(t^{+},t^{\prime+})&\Tilde{g}(t^{+},t^{\prime-})\\
        \Tilde{g}(t^{-},t^{\prime+})&\Tilde{g}(t^{-},t^{\prime-})
    \end{bmatrix}\;,
\end{align}
where $\pm$ refers to the positive and negative branch respectively. For causal theories, the entries of Eq. (\ref{eq:ContourGreensfunction}) are not independent. Indeed, if $t>t'$, one may traverse the track from $t$ to $\infty$ and back. Since this concerns only future times, by causality, no physical observable can change and hence $\Tilde{g}(t^{+},t^{\prime\pm}) = \Tilde{g}(t^{-},t^{\prime\pm})$. By an entirely similar argument, if $t<t'$, we find $\Tilde{g}(t^{\pm},t^{\prime +}) = \Tilde{g}(t^{\pm},t^{\prime-})$. 

To exploit this dependence, it is convenient to rotate the Keldysh space by
\begin{align}
    g = L\rho_{3}\Tilde{g}L^{\dagger},
\end{align}
where $L = 1/\sqrt{2}(1-i\rho_{2})$ and $\rho_{i},i = 1,2,3$ denote the Pauli matrices in Keldysh space. In this space, the Green's function takes the form
\begin{align}
    g = \begin{bmatrix}
        g^{R}&g^{K}\\0&g^{A}
    \end{bmatrix}\;,
\end{align}
where $g^{R,A,K}$ are the retarded, advanced and Keldysh Green's functions respectively. The retarded and advanced Green's functions describe equilibrium properties, such as the density of states. The Keldysh Green's function may be used to describe nonequilibrium physics and may be parametrized as $g^{K} = g^{R}F-Fg^{A}$, where $F$ represents the distribution function of the system. Here $F$ is diagonal in Nambu space, but may have any structure in spin space.

The Green's function  satisfies the Usadel equation, the saddle point equation of the NLSM. In the absence of spin-Hall and spin-swapping terms these correspond to Eqs. 7-9 in the main text. In the presence of spin-Hall and spin-swapping terms, current and torque are modified to
\begin{align}
    \label{eq:J-SGESH}
   \mathcal{J}_k &= -Dg\partial_kg +i\alpha_{kj}g[\sigma_{k},g] + \frac{i}{16}\varepsilon_{ijl}\gamma_{lk}
    \{[\sigma_i,g],\sigma_j + g\sigma_{j}g\}+i\frac{D}{4}\varkappa_{il}\varepsilon_{jkl}[[\sigma_{i},\partial_{j}g],g]+\frac{D}{4}\varepsilon_{jkl}\theta_{il}\{\sigma_{i}+g\sigma_{i}g,\partial_{j}g\}, \\
    \label{eq:T-SGESH}
    \mathcal{T} &= -i\alpha_{kj}[\sigma_{k},g\partial_{j}g]-\frac{i}{8}\epsilon_{ijl}\gamma_{lk} [\{\partial_{k}g,g\sigma_{i}g\},\sigma_j]-i\frac{D}{4}\varkappa_{il}\varepsilon_{jkl}\partial_{k}([[\sigma_{i},\partial_{j}g],g])-\frac{D}{4}\varepsilon_{jkl}\theta_{il}[\sigma_{i},g\partial_{j}g\partial_{k}g] \,.
\end{align}
%Let us analyze them  here for a  normal system. 
In this case, pair amplitudes are absent while the density of states is constant within the quasiclassical formalism. Thus, the retarded and advanced Green's function are equal to $\tau_{3}\delta(t-t')$ and $-\tau_{3}\delta(t-t')$ respectively, independent of spatial coordinates.

However, the Keldysh equation remains nontrivial, because the distribution functions may depend on space,  as happens when we apply a voltage. By substitution of $g^{R} = -g^{A} = \tau_{3}\delta(t-t')$  in these equations, we obtain a linear equation for $g^{K}$:
\begin{align}
    \partial_{k}J_{k}^{K} &= -\tau_{3}\partial_{t}g^{K}-i\alpha_{kj}\tau_{3}[\sigma_{j},\partial_{k}g^{K}]-\frac{i}{4}\gamma_{lk}(\epsilon_{ijl}\sigma_{i}\partial_{k}g^{K}\sigma_{j}+i\{\sigma_{l},g^{K}\}))\;,\\
   J^{K}_{k} &= -D\tau_{3}\partial_{k}g^{K}+i\alpha_{kj}\tau_{3}[\sigma_{j},g^{K}]+\frac{i}{4}\gamma_{lk}(\epsilon_{ijl}\sigma_{i}g^{K}\sigma_{j}-i\{\sigma_{l},g^{K}\})-i\frac{D}{2}\varkappa_{il}\varepsilon_{jkl}\tau_{3}[\sigma_{i},\partial_{j}g^{K}]+\frac{D}{2}\varepsilon_{jkl}\theta_{il}\{\sigma_{i},\partial_{j}g^{K}\}\;.
\end{align}
These equations may also be expressed in terms of the distribution functions, which in the normal state are related to the Keldysh component of the Green's function via $g^{K} = 2\tau_{3}F$, as:
\begin{align}
    \partial_{k}J_{k}^{K} &= -2\partial_{t}F-2i\alpha_{kj}[\sigma_{j},\partial_{k}F]-\frac{i}{2}\gamma_{lk}\tau_{3}(\epsilon_{ijl}\sigma_{i}\partial_{k}F\sigma_{j}+i\{\sigma_{l},F\}))\;,\\
   J^{K}_{k} &= -2D\partial_{k}F+i\alpha_{kj}[\sigma_{j},F]+\frac{i}{2}\gamma_{lk}\tau_{3}(\epsilon_{ijl}\sigma_{i}F\sigma_{j}-i\{\sigma_{l},F\})-iD\varkappa_{il}\varepsilon_{jkl}[\sigma_{i},\partial_{j}F]+D\varepsilon_{jkl}\theta_{il}\tau_{3}\{\sigma_{i},\partial_{j}F\}\;.
\end{align}
From these equations the diffusion equation for the physical observables, the chemical potential and excess spin can, be found via integration  over energy. 
We define the chemical potential $\mu =-\frac{1}{16}\text{Tr}g^{K}= -\frac{1}{32}\text{Tr}\tau_{3}\int dE F(E)$ and excess spin $S^{l}= -\frac{1}{16}\text{Tr}\tau_{3}\sigma_{l}g^{K} = -\frac{1}{32}\text{Tr}\sigma_{l}\int dE F(E)-F_{eq}(E)$ \cite{bergeret2018colloquium}, where $F_{eq}(E)$ is the equilibrium matrix distribution function governed by the Fermi-Dirac distribution. We express them in terms of the electrical current $j_{k} = \text{Tr}\tau_{3}J_{k}$  and spin current $j_{kl} = \text{Tr}\sigma_{l}J_{k}$. The resulting equations in the normal state are:
\begin{align}
    \partial_{k}j_{k} &=-\partial_{t}\mu\label{eq:Diffusionmu}\;,\\
    \partial_{k}j_{kl}&=-\partial_{t}S^{l}+\epsilon_{lmr} h_{m}S^{r}+\gamma_{lm}\partial_{m}\mu-\frac{1}{2}\Gamma_{0}S^{l}+\frac{1}{2}\Gamma_{lm}S^{m}+\alpha_{km}\epsilon_{lmr}\partial_{k}S^{r}\label{eq:DiffusionS}\;,\\
     j_{k}&=-D(\partial_{k}\mu-\theta_{il}\varepsilon_{jkl}\partial_{j}S^{i})+\gamma_{ik}S^{i}\label{eq:ElecCurr}\;,\\
    j_{kl} &= -D(\partial_{k}S^{l}-\theta_{il}\varepsilon_{jkl}\partial_{j}\mu-\varkappa_{ir}\varepsilon_{jkr}\varepsilon_{iml}\partial_{j}S^{m})-\alpha_{km}\epsilon_{lmr}S^{r}\label{eq:SpCurr}\;,
\end{align}
where $\Gamma_{0} = \text{Tr}\Gamma_{ab}$. These equations can be supplemented by suitable boundary conditions that depend on the type of boundary chosen, for example for a boundary with vacuum $n_{k}J_{k} = 0$ and $n_{k}J_{kl} = 0$ for all $l$.

The first two equations are the diffusion equations for charge and spin, respectively. As expected, charge conservation leads to a  zero  torque in Eq. (\ref{eq:Diffusionmu}).  For the spin current, there are several types of torques, see Eq. (\ref{eq:DiffusionS}). First of all, even in the absence of spin-charge coupling, there may be the torque of an effective exchange field.  $\gamma_{lm}\partial_{m}\mu$ is a generalization of the spin generation torque \cite{Gorini2017PRB}, it creates spin accumulation from electrical currents. On the other hand $-\frac{1}{2}\Gamma_{0}S^{l}+\frac{1}{2}\Gamma_{lm}S^{m}$ is a general spin relaxation term. For isotropic systems, $\Gamma_{lm} = \frac{1}{3}\Gamma_{0}\delta_{lm}$, and the torque reduces to $-\frac{1}{3}\Gamma_{0}S^{l}$. For anisotropic systems, spin-relaxation along one spin-axis may be different from that for other spin axes. Lastly, there is a torque due to spin precession. The spin-swapping and spin Hall terms do not give any torques in the normal state, even though their contribution to the torque is nonzero in the nonlinear Usadel equation. 

Next to this, both the electrical current and spin current definitions are altered. The electrical current, shown in Eq. (\ref{eq:ElecCurr}) consists of three terms. First of all there is the usual diffusion term $-D\partial_{k}\mu$. Then there is the spin-Hall contribution $\theta_{jki}\partial_{i}S^{j}$, which is responsible for the inverse spin-Hall effect. Because $\theta_{jki} = \epsilon_{jkl}\theta_{il}$ is necessarily antisymmetric in $i,k$, the generated electrical current is always perpendicular to the gradient of the spin. For isotropic materials $\theta_{jki} = \theta\varepsilon_{jki}$ and hence the spin of the generated spin current is perpendicular to both the current direction and spin gradient. However, for certain crystal structures the spin may as well be parallel to either of these directions as well. Lastly, there is the term $\gamma_{ik}S^{i}$,  gives a finite contribution to the electrical current in the presence of an excess spin and is therefore responsible for spin-galvanic effects.

The spin current, Eq. (\ref{eq:SpCurr}), has four contributions. Next to the usual diffusion term $-D\partial_{k}S^{l}$, there is the spin-Hall term $-\theta_{lkj}\partial_{j}\mu$, which generates a spin current from a gradient in the chemical potential, i.e. it is responsible for the spin-Hall effect. By structure of the spin-Hall tensor, the spatial direction of this spin current is perpendicular to the gradient. The allowed directions of spin depends on the symmetries of the material. For isotropic materials, this spin is required to be perpendicular to both the gradient and spin current directions. Lastly, there exists a generalization of the spin swapping term \cite{lifshits2009swapping} $\varkappa_{ijk}\epsilon_{ilm}\partial_{j}S^{m}$, which mixes spin currents that have both perpendicular spin and perpendicular direction. The allowed form of the tensor, and hence the allowed relative orientation between direction of the spin and spatial direction of the spin current depends on the crystal symmetry. Specifically, for isotropic materials $\varkappa_{ijk} = \varkappa\varepsilon_{ijk}$. Lastly, the spin-precession also alters the spin current.

The obtained spin diffusion equations contain all symmetry allowed terms, and are therefore completely general for normal metals with spin-charge coupling. The symmetry arguments only restrict the forms of the tensors, but do not relate the tensors of different terms. For this reason, the magnitudes of spin-precession, spin-relaxation and spin-generation are unrelated. For specific models, such as the SU(2) Rashba-like model \cite{virtanen2022nonlinear} the different tensors are related. However, one should take into account that extrinsic spin-orbit coupling is always allowed and alters these relations.
\newpage
\section{Vortex in a gyrotropic superconductor}\label{sec:Vortex}
In this section we study the excess spin around a vortex in a gyrotropic superconductor. For a vortex, the pair amplitudes vanish at the center ($r = 0$), while the phase of the pair amplitudes winds as $e^{in\phi}$ around the center of the vortex, where $\phi$ denotes the tangential coordinate and $n$ is the winding number of the vortex. Such vortices can be created for example in a type II superconductor by applying a magnetic field. %If the magnetic field is subsequently turned off, the vortices persist for a finite time, since they can only disappear via annihilation with a vortex with opposite winding or via the boundaries.

Now, according to Eq. (\textcolor{purple}{9}) in the main text, a gradient in the Green's function is coupled to the spin via the term in the torque $\gamma_{kj}\sigma_{j}\partial_{k}g_{s}$, where $g_{s}$ is the Green's function in the absence of a spin-galvanic effect. We may distinguish two different kinds of spatial dependence, spatial dependence of the phase and spatial dependence of the magnitude. Indeed, for a general system, we may write \begin{align}
    g_{s}(\vec{r}) = e^{-i\phi(\vec{r})\tau_{3}/2}(g_{s0}(\vec{r})\tau_{3}+f_{s0}(\vec{r})\tau_{2})e^{i\phi(\vec{r})\tau_{3}/2}\;,
\end{align}
If the phase depends on position, the spatial derivative has opposite sign for the $\langle\psi\psi\rangle$ and $\langle\psi^{\dagger}\psi^{\dagger}\rangle$ correlations, and it appears as $\nabla\phi[\tau_{3},g]$. Thus, in the presence of a spin-galvanic term the Usadel equation has  an additional term $\partial_{k}\phi\gamma_{lk}[\sigma_{l}\tau_{3},g]$. This term appears exactly like an exchange field.

On the other hand, if the magnitude of the pair amplitudes depends on position, there is no time-reversal symmetry breaking, and hence it is not like an exchange field, and it can not introduce any spin dependence in the density of states. %Indeed, considering that the normalization condition needs to be satisfied, but the phase is not affected, such terms are proportional to $[\sigma_{l}e^{i\phi\tau_{3}}\tau_{2},g]$, and we conclude it is merely a source of triplet correlations. {\color{red} SB: this is not clear to me.}\textcolor{blue}{TK: it appears as $\tau_{2}\sigma_{l}$, that is, in the same way as a triplet pair potential.}

In the specific case of a vortex, the effective exchange field can be written as $n\gamma_{kj}\vec{\hat{\phi}}_{k}[\tau_{3}\sigma_{j},g_{s}]$. Thus, it takes the same form an an exchange field, where the spin direction is $\vec{\hat{\phi}}_{k}\gamma_{kj} = \cos\phi(\gamma_{xx},\gamma_{xy},0)+\sin\phi(\gamma_{yx},\gamma_{yy},0)$ and $\gamma_{xx,xy,yx,yy}$ are coefficients independent of the angle. Consequently, the direction of the effective exchange field is given by 
\begin{align}
\vec{h}_{\text{eff},j} = n\vec{\hat{\phi}}_{k}\gamma_{kj}\;.
\end{align}

To derive the symmetry of the spin for generic forms of the spin-galvanic tensor, it is instructive to consider three cases separately, in all of which the tensor has two nonvanishing elements. First of all, we consider $\gamma_{xy} = -\gamma_{yx}$. A well-known example of spin-orbit coupling that leads to a spin-galvanic tensor of this form is Rashba spin orbit coupling. Generally it is the only allowed symmetry in the $C_{3v},C_{4v}$ and $C_{6v}$ point groups. For this symmetry the spin is always perpendicular to the direction of the current and hence around a vortex the spin points in the radial direction. The sign of $n\gamma_{xy}$ determines whether this is radially outward ($>0$) or inward ($<0$). Moreover, the magnitude of the excess spin is homogeneous at a fixed distance from the vortex core, it is independent of $\phi$.

Next to this we may consider the chiral symmetry. The chiral symmetry is defined by having $\gamma_{xx} = \gamma_{yy}$ as the only nonzero elements of the spin-galvanic tensor. Second rank pseudotensors are required to have this symmetry in the $T,O,D_{3,4,6}$ groups. In this case the spin-galvanic tensor is proportional to the identity tensor and hence the spin always points parallel or antiparallel to the current. Around the vortex this implies the spin points in the tangential direction. 

Lastly, we may consider the case $\gamma_{xx} = -\gamma_{yy}$. This symmetry is equivalent to $\gamma_{xy} = \gamma_{yx}$ as can be seen by rotating the definition of the axes by $\pi/4$ from $(x,y)$ to $(\frac{x+y}{\sqrt{2}},(\frac{y-x}{\sqrt{2}}))$. Such terms arise for example for Dresselhaus spin-orbit coupling and are the only allowed ones in the point groups $S_{4},D_{2d}$. Just like in the two other cases, the spin rotates once when going around the vortex, but unlike for the other two cases, if one traverses around the vortex clockwise the spin does not rotate clockwise but rather counterclockwise, i.e. it is given by $e^{-i\phi\sigma_{z}}\sigma_{x,y}$ instead of $e^{i\phi\sigma_{z}}\sigma_{x,y}$. Hence, in this case the spin is not always along the radial or tangential direction, the angle between current direction and spin direction changes continuously. 

\begin{figure}
    \centering
    \includegraphics[width = 8.6cm]{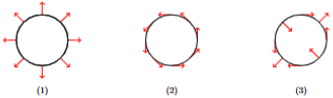}
    \caption{The direction of spin around the vortex for three different symmetries. \textit{(a):} $\gamma_{xy} = -\gamma_{yx}$, as allowed in point groups $C_{3v,4v,6v}$. \textit{(b):} $\gamma_{xx} = \gamma_{yy}$, as allowed in point groups $T,O,D_{4}$. \textit{(c):} $\gamma_{xx} = -\gamma_{yy}$, as allowed in point groups $S_{4},D_{2d}$.}
    \label{fig:SymmetrySpin}
\end{figure}

The spin texture around the vortex is illustrated for each of these three different symmetries in Fig. \ref{fig:SymmetrySpin}. For any other gyrotropic group, a combination of (a subset of) the above three types of gyrotropic tensors are allowed, and hence to first order in the spin-galvanic tensor the spin texture is a weighted sum of the previous three cases. Thus, a combination of Rashba-like and chiral symmetry groups, as is the only allowed combination in materials with $C_{3,4,5}$ symmetry, there is a fixed angle between current direction and direction of the spin, which generally is neither $0$ nor $\pi/2$. Meanwhile for the symmetry group $C_{1v,2v}$, in which the only allowed terms in the gyrotropic tensor are the Rashba-like and Dresselhaus-like terms, or in the symmetry group $D_{2}$, in which the gyrotropic tensor contains terms with chiral and Dresselhaus-like symmetries, the angle between the current and spin direction is not constant, but varies around the average given by $\pi/2$ or $0$ respectively. Lastly, in the $C_{1,2}$ group all terms are allowed and the angle between current and spin direction varies around an average that does not need to be $0$ or $\pi/2$.

Next, we consider the dependence of the excess spin on the radial coordinate. Since the chiral and Rashba-like symmetries can be obtained from each other by a homogeneous spin rotation, while the Dresselhaus-like symmetries can be found from the other two by changing $\vec{\hat{\phi}}\mapsto-\vec{\hat{\phi}}$, we conclude that the radial dependence of the excess spin is the same for all of these cases. Moreover, since any gyrotropic tensor can be written as the linear combination of these different cases, the spatial dependence is the same for any tensor $\gamma_{kc}$. For this reason for explicit calculation we focus solely on the case with Rashba-like symmetry.

We now compute explicitly the current  and excess spin densities by solving the Usadel equation, Eq. (\textcolor{purple}{7-9}) in the main text with $\alpha_{kj} = 0$, which read
\begin{align}
\label{eq:Usadel-SGE}
\partial_k\mathcal{J}_k+ [\tau_3(\hat{\omega} + i\mathbf{h}{\bm\sigma})+\hat{\Delta},g] = \mathcal{T} - \frac{1}{8}\Gamma_{jk}[\sigma_jg\sigma_k,g] \; ,   
\end{align}
where the matrix current $\mathcal{J}_k$ and matrix torque $\mathcal{T}$ are:
\begin{align}
    \label{eq:J-SGE}
    \mathcal{J}_k &= -Dg\tilde{\nabla}_kg + \frac{i}{16}\epsilon_{ijl}\gamma_{lk}
    \{[\sigma_i,g],\sigma_j + g\sigma_jg\}\;, \\
    \label{eq:T-SGE}
    \mathcal{T} &= -\frac{i}{8}\epsilon_{ijl}\gamma_{lk} [\{\partial_kg,g\sigma_ig\},\sigma_j] \;.
\end{align}

We assume an isotropic spin-relaxation $\Gamma_{jk} = \frac{D}{l_{\text{so}^{2}}}\delta_{jk}$, where $l_{\text{so}}$ is the spin-relaxation length, for simplicity of notation. 
For this we use the  Riccati parameterization (we use the notation $a,b$ instead of the usual $\gamma, \Tilde{\gamma}$ to avoid confusion with the tensor of the gyrotropic term) \begin{align}g = \begin{bmatrix}
    (1+ab)^{-1}(1-ab)\text{sign}(\omega_{n})&2(1+ab)^{-1}a\\
    2(1+ba)^{-1}b&-(1+ba)^{-1}(1-ba)\text{sign}(\omega_{n})
\end{bmatrix}\; . \end{align}
This  parametrization guarantees the normalization condition $g^2=1$. %The equations can be found by substituting the Riccati parameterization in a term $A$ and then calculating the contribution to the equation for $\nabla^{2}a$ via  $\frac{1}{4}(\mathbf{1}+ab)\Big(\text{sign}(\omega)\text{Tr}\sigma_{d}(\tau_{1}+i\tau_{2})A-\text{Tr}\sigma_{d}(\mathbf{1}+\tau_{3})A a\Big)$, where $d = 0,1,2,3$, while the contribution to the equation for $b$ can be found via $\frac{1}{4}(\mathbf{1}+ba)\Big(-\text{sign}(\omega)\text{Tr}\sigma_{d}(\tau_{1}-i\tau_{2})A-\text{Tr}\sigma_{d}(\mathbf{1}-\tau_{3})Ab\Big)$.
For a single vortex, in the absence of a gyrotropic tensor ($\gamma_{lk} = 0$) the equation for $a$ can be written as 
\begin{align}
    \nabla^{2}a -a\nabla a\cdot\nabla b = 2\frac{|\omega|}{D}a + \frac{2\Delta}{D} e^{in\phi} (1-a^2-( be^{in\phi}-a e^{-in\phi}))+\frac{1}{l_{\text{so}}^{2}}(a-\sigma_{k} a\sigma_{k})\;,    
\end{align}
while the equation for $b$ is found similarly.
In polar coordinates this equation can be written as
\begin{align}
    \frac{\partial^{2}a}{\partial r^{2}}+\frac{1}{r}\frac{\partial a}{\partial r}+\frac{1}{r^{2}}\frac{\partial^{2}a}{\partial \phi^{2}}- a\Big(\frac{\partial a}{\partial r}\frac{\partial b}{\partial r}+\frac{1}{r^{2}}\frac{\partial a}{\partial\phi}\frac{\partial b}{\partial\phi}\Big) = \frac{2|\omega|}{D}a-\frac{\Delta}{D}(1-a^{2}-( be^{in\phi}-a e^{-in\phi}))+\frac{1}{l_{\text{so}}^{2}}(a-\sigma_{k}a\sigma_{k})\;.
\end{align}
This is a differential equation in two variables, $r$ and $\phi$. 
However, as discussed before, by symmetry the magnitude of the pair potential depends only on the radial coordinate, while its phase depends only on the tangential coordinate. Moreover, in the absence of a gyrotropic tensor there are no triplet correlations. We may write
\begin{align}
    a_0(r,\phi) =  a_{s}(r)e^{in\phi}\;,\\
     b_0(r,\phi) =  a_{s}(r)e^{-in\phi}\;.
\end{align}
The equation for $a_{s}(r)$ is a second order nonlinear equation with nonconstant coefficients in one variable;
\begin{align}\label{eq:aRadial}
    \frac{\partial^{2} a_{s}}{\partial r^{2}}+\frac{1}{r}\frac{\partial  a_{s}}{\partial r}-\frac{n^{2}}{r^{2}} a_{s}- a_{s}\Big(\left(\frac{\partial a_{s}}{\partial r}\right)^{2}+\frac{n^{2}}{r^{2}} a_{s}^{2}\Big) = \frac{2|\omega|}{D} a_{s}-\frac{\Delta}{D}(1- a_{s}^{2})\;.
\end{align}
We may equip it with the boundary conditions $ a_{s}(r = 0) = 0$ and $\lim_{r\xrightarrow{}\infty}\frac{\partial  a_{s}}{\partial r} = 0$. This system of equations can be solved using \MATLAB solver bvp5c.

From the solution the current density flowing around the vortex can be computed via
\begin{align}
    I(r) = \frac{-i}{1+a_{0} b_{0}}(a_{0}\nabla b_{0}- b_{0}\nabla a_{0} ) = \frac{1}{r}\frac{n a_{s}^{2}}{1+ a_{s}^{2}}\vec{\hat{\phi}}\;.
\end{align} 
We note that while there exists a factor $\frac{1}{r}$ in the expression, the current density does not diverge near the center, since $ a_{s}$ has to be at least first order in $r$ following Eq. \eqref{eq:aRadial}. Hence, the current density vanishes at $r = 0$.

Next we may consider the inclusion of a gyrotropic tensor. To first order we may write $a = a_{0}+\delta g_{t}$ and similarly for $ b$, where $a_{0} =  a_{s}e^{in\phi}$ and $ b_{0} =  a_{s}e^{-in\phi}$. 

For all terms except the spin-galvanic one we may write $ a\approx a_{s}+\delta a_{t}$ and expand to first order in $\delta a_{t}$, since this term, as shown above, is absent if there is no spin galvanic effect. For the spin-galvanic term we only take the zeroth order in $a$, i.e. $ a_{s}e^{in\phi}$. This term reads $\gamma_{xl}\partial_{x} a_{s}e^{in\phi}\sigma_{l}+\gamma_{yl}\partial_{y} a_{s}e^{in\phi}\sigma_{l} = \sigma_{l}\left(\gamma_{xl}(\cos\phi\frac{\partial a_{s}}{\partial r}-\frac{in\sin\phi}{r}  a_{s})+\gamma_{yl}(\sin\phi\frac{\partial a_{s}}{\partial r}+\frac{in\cos\phi}{r} a_{s})\right) = \sigma_{c}\left((\gamma_{xl}\cos{\phi}+\gamma_{yl}\sin\phi)\frac{\partial  a_{s}}{\partial r}+\frac{in}{r}(-\gamma_{xl}\sin{\phi}+\gamma_{yl}\cos{\phi})\right)$.

Denoting $\gamma_{xy} = -\gamma_{yx} = -\gamma_{o}$ we obtain for the radial derivative contribution $-\cos{\phi}\sigma_{y}+\sin\phi\sigma_{x}\frac{\partial a_{s}}{\partial r} = -\sigma_{y}e^{i\phi\sigma_{z}}\frac{\partial a_{s}}{\partial r}$ and for the tangential derivative contribution
$\frac{in}{r}(\cos\phi\sigma_{x}+\sin\phi\sigma_{y})  = \frac{in}{r}\sigma_{x}e^{i\phi\sigma_{z}} a_{s}$. 

The equation for $\delta a_{t}$ can be found as
\begin{align}
    &\frac{\partial^{2}\delta a_{t}}{\partial r^{2}}+\frac{1}{r}\frac{\partial\delta a_{t}}{\partial r}+\frac{1}{r^{2}}\frac{\partial^{2}\delta a_{t}}{\partial \phi^{2}}-\delta a_{t}(\frac{\partial a_{s}}{\partial r })^{2}- a_{s}\frac{\partial a_{s}}{\partial r}\frac{\partial\delta a_{t}}{\partial r}-e^{2ni\phi} a_{s}\frac{\partial a_{s}}{\partial r}\frac{\partial\delta b_{t}}{\partial r}-\frac{n^{2}}{r^{2}} a_{s}^{2}\delta a_{t}
    +\frac{in}{r^{2}} a_{s}^{2}\frac{\partial\delta a_{t}}{\partial \phi}- \frac{in}{r^{2}}e^{2ni\phi} a_{s}^{2}\frac{\partial\delta b_{t}}{\partial \phi} \nonumber\\&   = \frac{2|\omega|}{D}\delta a_{t}+\frac{2\Delta}{D} a_{s}\delta b_{t}+\frac{\Delta}{D}(\delta b_{t}e^{in\phi}-\delta a_{t}e^{-in\phi})+\gamma_{o}(\frac{in}{r} a_{s}\sigma_{x}e^{i\phi\sigma_{z}}-\frac{\partial  a_{s}}{\partial r}\sigma_{y}e^{i\phi\sigma_{z}})e^{in\phi}\; ,\\
    &\frac{\partial^{2}\delta b_{t}}{\partial r^{2}}+\frac{1}{r}\frac{\partial\delta b_{t}}{\partial r}+\frac{1}{r^{2}}\frac{\partial^{2}\delta b_{t}}{\partial \phi^{2}}-\delta b_{t}(\frac{\partial^{2} a_{s}}{\partial r })^{2}- a_{s}\frac{\partial a_{s}}{\partial r}\frac{\partial\delta b_{t}}{\partial r}- a_{s}\frac{\partial a_{s}}{\partial r}\frac{\partial\delta a_{t}}{\partial r}e^{-2in\phi}-\frac{n^{2}}{r^{2}} a_{s}^{2}\delta a_{t}
    -\frac{in}{r^{2}} a_{s}^{2}\frac{\partial\delta b_{t}}{\partial \phi}+\frac{in}{r^{2}} a_{s}^{2}\frac{\partial\delta a_{t}}{\partial \phi} e^{-2ni\phi}    \nonumber\\&= \frac{2|\omega|}{D}\delta b_{t}+\frac{\Delta}{D}(2 a_{s}\delta a_{t}-(\delta b_{t}e^{in\phi}-\delta a_{t}e^{-in\phi}))-\gamma_{o}(-\frac{in}{r} a_{s}\sigma_{x}e^{i\phi\sigma_{z}}+\frac{\partial  a_{s}}{\partial r}\sigma_{y}e^{i\phi\sigma_{z}})e^{-in\phi}\; .
\end{align}
We note that these equations predict the absence of any excess spin for $n = 0$, as required by the absence of any current in this case. Indeed, for $n = 0$ we find $ a_{t} = - b_{t}$ and $\alpha = \Tilde{\alpha}$ as solutions to the problem and hence $ a_{s}(\delta a_{t}+\delta b_{t}) = 0$. As soon as $n\neq 0$ we have $ a_{t}\neq  b_{t}$ and a nonzero spin is generated.

Again exploiting the symmetry of the problem we may write $\delta  a_{t} (r,\phi)= ( a_{x}(r)\sigma_{x}+ a_{y}(r)\sigma_{y})e^{i\phi \sigma_{z}}e^{in\phi}$ and $\delta  b_{t}(r,\phi) = \sigma_{y}\delta a_{t}(r,\phi)^{*}\sigma_{y} = -( a_{x}(r)^{*}\sigma_{x}+ a_{y}(r)^{*}\sigma_{y})e^{i\phi\sigma_{z}}e^{-in\phi}$ to obtain two coupled nonlinear second order differential equations. We first consider the case in which the tensor of the gyrotropic term has the same symmetry as for Rashba spin orbit coupling, i.e. $\gamma_{xy} = -\gamma_{yx} = \gamma_{o}$, while all other terms of the tensor vanish. In this case
\begin{align}
    &\frac{\partial^{2} a_{x}}{\partial r^{2}}+\frac{1}{r}\frac{\partial  a_{x}}{\partial r}-\frac{n^{2}+1}{r^{2}} a_{x}-\frac{2in}{r^{2}} a_{y}-(\frac{\partial a_{s}}{\partial r})^{2} a_{x}-2i a_{s}\frac{\partial  a_{s}}{\partial r}\text{Im}\frac{\partial a_{x}}{\partial r} - 3\frac{n^{2}}{r^{2}} a_{s}^{2} a_{x}+2in a_{s}^{2}\text{Re} a_{y}\nonumber\\&= \frac{2|\omega|}{D} a_{x}+\frac{1}{l_{\text{so}}^{2}} a_{x}+\frac{2\Delta}{D} a_{s} a_{x}+\frac{2\Delta}{D}\text{Re} a_{x}+\frac{in}{r}\gamma_{o} a_{s}\; ,\\
    &\frac{\partial^{2} a_{y}}{\partial r^{2}}+\frac{1}{r}\frac{\partial  a_{y}}{\partial r}-\frac{n^{2}+1}{r^{2}} a_{y}+\frac{2in}{r^{2}} a_{x}-(\frac{\partial a_{s}}{\partial r})^{2} a_{y}-2i a_{s}\frac{\partial  a_{s}}{\partial r}\text{Im}\frac{\partial a_{y}}{\partial r} - 3\frac{n^{2}}{r^{2}} a_{s}^{2} a_{x}-2in a_{s}^{2}\text{Re} a_{x}\nonumber\\&= \frac{2|\omega|}{D} a_{y}+\frac{1}{l_{\text{so}}^{2}} a_{y}+\frac{2\Delta}{D} a_{s} a_{y}+\frac{2\Delta}{D}\text{Re} a_{y}-\gamma_{o}\frac{\partial a_{s}}{\partial_{r}}\; .
\end{align}
On the Matsubara track, since $ a_{s}$ is real, we may require $ a_{y} = a_{2}$ to be real and $ a_{x} = ia_{1}$ to be imaginary. The equations for the real variables $a_{1,2}$ are
\begin{align}
    &\frac{\partial^{2}a_{1}}{\partial r^{2}}+\frac{1}{r}\frac{\partial a_{1}}{\partial r}-\frac{n^{2}+1}{r^{2}}a_{1}-\frac{2n}{r^{2}}a_{2}-(\frac{\partial a_{s}}{\partial r})^{2}a_{1}-2 a_{s}\frac{\partial a_{s}}{\partial r}\frac{\partial a_{1}}{\partial r} - 3\frac{n^{2}}{r^{2}} a_{s}^{2}a_{1}+2n a_{s}^{2}a_{2}\nonumber\\&= \frac{2|\omega|}{D}a_{1}+\frac{1}{l_{\text{so}}^{2}}a_{1}+\frac{2\Delta}{D} a_{s}a_{1}+\frac{n}{r}\gamma_{o} a_{s}\; ,\\ 
    &\frac{\partial^{2}a_{2}}{\partial r^{2}}+\frac{1}{r}\frac{\partial a_{2}}{\partial r}-\frac{n^{2}+1}{r^{2}}a_{2}-\frac{2n}{r^{2}}a_{1}-(\frac{\partial a_{s}}{\partial r})^{2}a_{2} - 3\frac{n^{2}}{r^{2}} a_{s}^{2}a_{2}\nonumber\\&= \frac{2|\omega|}{D}a_{2}+\frac{1}{l_{\text{so}}^{2}}a_{2}+\frac{2\Delta}{D}(1+ a_{s})a_{2}-\gamma_{o}\frac{\partial a_{s}}{\partial_{r}}\; .
\end{align}
The spin dependent part of the density of states may now be written as
$-2\text{sign}(\omega) a_{s}(\delta a_{t}e^{-in\phi}+\delta b_{t}e^{in\phi}) = -4\text{sign}(\omega) a_{s}\text{Im}( a_{x}\sigma_{x}+ a_{y}\sigma_{y})e^{i\phi}\sigma_{z} = -4\text{sign}(\omega) a_{s}a_{1}\sigma_{x}e^{i\phi\sigma_{z}}$. We note that this function is always even in $\omega$, since $ a_{s}(\omega) =  a_{s}(-\omega)$ but $a_{1}(\omega) = -a_{1}(-\omega)$.

We confirm that for $n = 0$ nonetheless $ a_{s}$ and $a_{2}$ are the only nonvanishing terms, and hence the density of state has no spin-dependence. For $n\neq 0$ this is not the case and an excess spin is generated.

We numerically studied the case $n = 1$. We may consider two different limits, $T\ll \Delta(T)$ and $T\gg \Delta(T)$. Specifically, we chose $T/T_{c} = 0.05$, for which $\Delta$ is suppressed to $\Delta(T)\approx\Delta(T = 0) = \Delta_{0}$, and $T = 0.98 T_{c}$, for which $\Delta\approx 0.243\Delta_{0}$. We decided to ignore self-consistency of the pair potential around the vortex in the calculation and take into account only the temperature dependence, since it was found that the spatial dependence of the pair amplitude provides only small corrections to the pair amplitude. In our calculations we choose a relatively strong spin-relaxation, $\Delta_{0}\tau_{\text{so}} = 0.1$.

\begin{figure}
    \centering
    \includegraphics[width = 8.6cm]{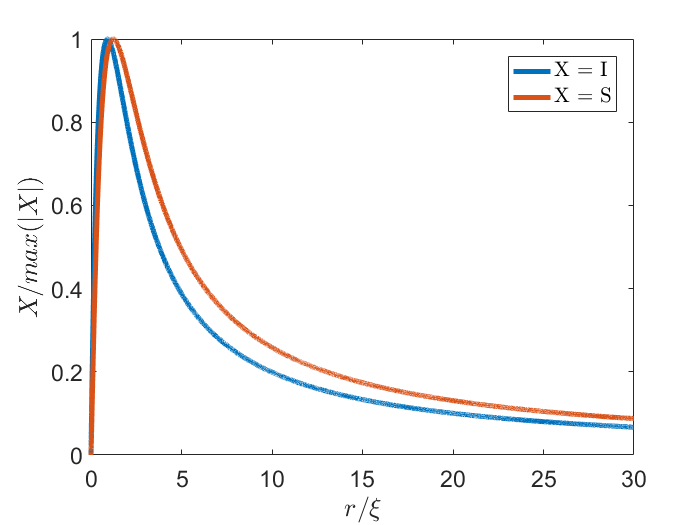}
    \includegraphics[width = 8.6cm]{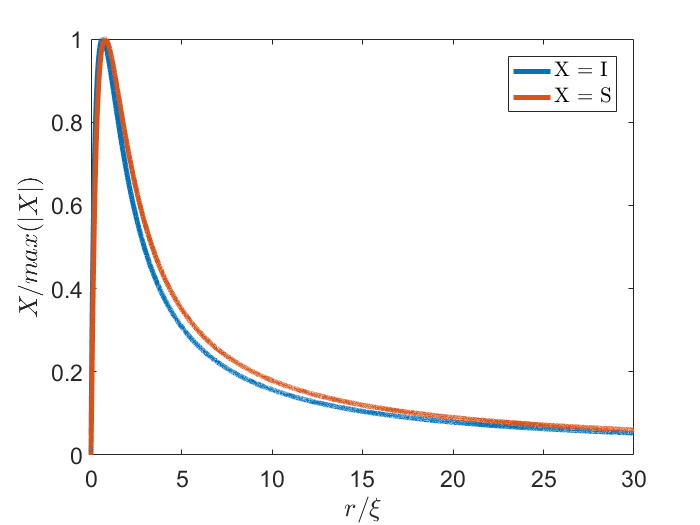}
    \caption{The current density ($I$) and excess spin density ($S$) around a vortex with $n = 1$ normalized to their maximum values as a function of the radial coordinate, for (left) $T/T_{c} = 0.05$ and (right) $T/T_{c} = 0.98$. The maximum in the excess spin density is further away from the vortex core than the maximum of the current density. The difference becomes smaller as the temperature approaches the critical temperature.}
    \label{fig:smallTandlargeTSM}
\end{figure}

The results are shown in Fig. \ref{fig:smallTandlargeTSM}, of which the left panel corresponds to the figure in the main text. We show the current density $I$ normalized by the maximum current density and the excess spin density $S^{l}$ normalized by the maximum, as a function of the radial coordinate, which is normalized by $\xi = \sqrt{D/2\Delta(T)}$. Clearly, for low temperatures, the spatial dependence of the excess spin density differs from the spatial dependence of the current density, with the spin density profile having its maximum further away from the vortex core. This is in contrast with the GL theory prediction, in which the spin depends linearly on the local current. For larger temperatures the difference becomes smaller. However, we notice that even for $T/T_{c} = 0.98$, i.e. close to the critical current, there is still a noticeable difference between the two spatial profiles.

\newpage

\section{The Ginzburg-Landau limit: Derivation of the Lifshtiz-invariant}\label{sec:GL}
In this section we derive the Ginzburg-Landau (GL) free energy from the Usadel equation. The GL functional is a convenient description for superconductors at temperatures close to the critical temperature, so that $\Delta(T)\ll k_{B}T$.
Therefore, near the critical temperature, we may linearize the Usadel equation in pair amplitudes and use the solution combined with the self-consistency relation to obtain the GL functional as has been shown before for both the linearized Usadel and Eilenberger equations \cite{gor1959microscopic,golubov2003upper,ilic2022theory,houzet2015quasiclassical}. We will here follow an approach similar to the one presented in those articles. 
We approximate the quasiclassical Green's function as
\begin{align}
    g\approx \begin{bmatrix}
        1-\frac{\Delta\Tilde{\Delta}}{2}f_{0}\Tilde{f}_{0}&\Delta f_{0}+\Delta^{2}\Tilde{\Delta}f_{1}\\
\Tilde{\Delta }\Tilde{f}_{0}+\Delta\Tilde{\Delta}^{2}\Tilde{f}_{1}&    -1+\frac{\Delta\Tilde{\Delta}}{2}\Tilde{f}_{0}f_{0}
\end{bmatrix}\;,
\end{align}
where $f_{0,1}$ and $\Tilde{f}_{0,1}$ are functions independent of $\Delta,\Tilde{\Delta}$, which in real space are related by $\Tilde{f}_{0,1} = f_{0,1}^{*}$. 
In this limit, only the derivative term, the energy term, the exchange field, the spin-galvanic term, and the spin relaxation term remain in the equation. 
Indeed, we apply the Fourier transform and obtain
\begin{align}
    &(|\omega| + D_{ij}q_{i}q_{j})f_{0}-\Delta+i\text{sign}(\omega)(h_{k}+q_{l}\gamma_{kl})\{\sigma_{k},f_{0}\}-\Gamma_{ij}(f_{0}-\sigma_{i}f_{0}\sigma_{j}) = 0\; .
\end{align}
From this equation, it is clear that the spin-galvanic term appears only as an effective exchange field $q_{l}\gamma_{kl}$. In other words, it renormalizes the latter as $\tilde h_k= h_k+q_l\gamma_{kl}$. 
The term in the free energy due to an exchange or Zeeman field can be written as 
\begin{align}\label{eq:generalizedh}
    F_Z=\frac{1}{2}\chi_{ij}\tilde h_{i}\tilde h_{j}\; ,
\end{align}
where $\chi_{ij}$ is the response tensor for the excess spin. The Lifshitz-invariant is the term proportional to $q$ of the previous expression. Thus,
\begin{align}\label{eq:Lifschitzlike}
    F_L=\chi_{ij}h_j\gamma_{ik}\partial_k\varphi \; . 
\end{align}
This expression is identical to the expression for the Lifshitz invariant of the Ginzburg-Landau functional presented in \cite{AgterbergInBook2012} if we define the second rank pseudotensor as $K_{jk} = \chi_{ij}\gamma_{ik}/\Delta^{2}$. However, since a phase gradient appears in the same way as a magnetic field even in the nonlinear equation, the lowest order response to $h_{k}+q_{l}\gamma_{kl}$ for any temperature has the form of Eq. (\ref{eq:generalizedh}), with suitable definition of $\chi_{ij}(T)$. Therefore, Eq. (\ref{eq:Lifschitzlike}) is more generic than the expression for the Lifshitz invariant in \cite{AgterbergInBook2012}, it represents the contribution to the Free energy linear in the phase gradient for any temperature.

\end{document}